%% file: paper.tex
\documentclass[twocolumn]{emulateapj}
\usepackage[]{units}
\usepackage{datetime}
\usepackage{graphicx}
\usepackage{amssymb}
\usepackage{amsmath}
\usepackage{threeparttable}
\usepackage{url}
\usepackage{color}

\newcommand{\spi}{{\it Spitzer}}

\newcommand{\oviii}{SDSS~J0834$+$0159}
\newcommand{\oxii}{SDSS~J1232$+$0912}
\newcommand{\oxxii}{SDSS~J2215$-$0056}
\newcommand{\oxxiii}{SDSS~J2323$-$0100}
\mathchardef\mhyphen="2D
\shorttitle{Extreme outflows in quasars}
\shortauthors{Zakamska et al.}

\begin{document}

\title{Discovery of extreme [OIII]$\lambda$5007\AA\ outflows in high-redshift red quasars\footnote{Based on observations made with ESO Telescopes at the La Silla Paranal Observatory under program ID 093.B-05553.}}

\author{Nadia L. Zakamska\altaffilmark{1,2}, Fred Hamann\altaffilmark{3}, Isabelle P\^aris\altaffilmark{4}, W. N. Brandt\altaffilmark{5}, Jenny E. Greene\altaffilmark{6}, Michael A. Strauss\altaffilmark{6}, Carolin Villforth\altaffilmark{7,8}, Dominika Wylezalek\altaffilmark{1}, Rachael M. Alexandroff\altaffilmark{1}, Nicholas P. Ross\altaffilmark{9}}
\altaffiltext{1}{Department of Physics \& Astronomy, Johns Hopkins University, Bloomberg Center, 3400 N. Charles St., Baltimore, MD 21218, USA}
\altaffiltext{2}{Deborah Lunder and Alan Ezekowitz Founders' Circle Member, Institute for Advanced Study, Einstein Dr., Princeton, NJ 08540, USA}
\altaffiltext{3}{Department of Physics and Astronomy, University of California, Riverside, CA, 92507, USA}
\altaffiltext{4}{LAM - Laboratoire d'Astrophysique de Marseille, P\^ole de l\'Etoile Site de Ch\^ateau-Gombert 38, rue Fr\'ed\'eric Joliot-Curie 13388 Marseille cedex 13, France}
\altaffiltext{5}{Department of Astronomy \& Astrophysics, 525 Davey Lab, The Pennsylvania State University, University Park, PA 16802, USA; Institute for Gravitation and the Cosmos, The Pennsylvania State University, University Park, PA 16802, USA; Department of Physics, 104 Davey Lab, The Pennsylvania State University, University Park, PA 16802, USA}
\altaffiltext{6}{Department of Astrophysical Sciences, Princeton University, Princeton, NJ 08544, USA}
\altaffiltext{7}{Scottish Universities Physics Alliance (SUPA), University of St Andrews, School of Physics and Astronomy, North Haugh, KY16 9SS, St Andrews, Fife, UK}
\altaffiltext{8}{University of Bath, Department of Physics, Claverton Down, Bath, BA2 7AY, UK}
\altaffiltext{9}{Scottish Universities Physics Alliance (SUPA), Institute for Astronomy, University of Edinburgh, Royal Observatory, Edinburgh EH9 3HJ, UK}

\begin{abstract}
Black hole feedback is now a standard component of galaxy formation models. These models predict that the impact of black hole activity on its host galaxy likely peaked at $z=2-3$, the epoch of strongest star formation activity and black hole accretion activity in the Universe. We used XShooter on the Very Large Telescope to measure rest-frame optical spectra of four $z\sim 2.5$ extremely red quasars with infrared luminosities $\sim 10^{47}$ erg s$^{-1}$. We present the discovery of very broad (full width at half max$= 2600-5000$ km s$^{-1}$), strongly blue-shifted (by up to 1500 km s$^{-1}$) [OIII]$\lambda$5007\AA\ emission lines in these objects. In a large sample of type 2 and red quasars, [OIII] kinematics are positively correlated with infrared luminosity, and the four objects in our sample are on the extreme end both in [OIII] kinematics and infrared luminosity. We estimate that at least 3\% of the bolometric luminosity in these objects is being converted into the kinetic power of the observed wind. Photo-ionization estimates suggest that the [OIII] emission might be extended on a few kpc scales, which would suggest that the extreme outflow is affecting the entire host galaxy of the quasar. These sources may be the signposts of the most extreme form of quasar feedback at the peak epoch of galaxy formation, and may represent an active ``blow-out'' phase of quasar evolution. 
\end{abstract}


\section{Introduction}

While supermassive black holes have masses only $\sim 0.1\%$ of their host galaxies, they are now thought to exert a significant controlling effect on galaxy evolution \citep{tabo93, silk98, spri05}. The enormous power of the relativistic outflows and the radiation from the accretion disk may be the critical agent limiting the mass of galaxies in the Universe \citep{thou95, crot06}. At the same time, the strong correlations between black hole masses and the velocity dispersions of their hosts in the present-day universe \citep{gebh00, ferr00, gult09} and the similarity of the cosmic evolution of star formation and black hole accretion \citep{boyl98, hopk08, mada14} also suggest a connection between the formation of supermassive black holes and their host galaxies. 

One possibility for such a connection, known as ``quasar feedback'', is that the energy output of the black hole in its most luminous (quasar) phase becomes coupled to the gas from which stars in the host galaxy would otherwise form. The gas is then reheated or pushed out of the galactic potential \citep{hopk06}, resulting in suppression of star formation and in a ``quenched'' galaxy. Just like the winds driven by powerful starbursts \citep{heck90, veil94, veil05}, such galaxy-wide quasar-driven winds are likely to be inhomogeneous, with different phases of the wind medium observable in different domains of the electromagnetic spectrum. As a result, observational evidence for quasar-driven winds has been accumulating via multiple different observational techniques which trace different components of the wind (e.g., \citealt{arav08, char09, nesv10, ogle10, rupk13b, veil13a, tomb13, gree14a, sun14, cric15, nard15}).  

One fruitful technique is observations of forbidden emission lines of ionized gas. Because luminous quasars can easily ionize all the available gas over the entire galaxy and even into intergalactic space \citep{stoc87, liu09, vill11a, hain13}, the large spatial extents of ionized gas by themselves do not constitute evidence for quasar feedback or a wind: very extended narrow line emission could simply be due to gas left-over from a merger event or even a small companion galaxy illuminated by the quasar. 

Rather, a tell-tale sign of a galactic wind is the presence of strongly kinematically disturbed gas at large distances from the galactic center, as measured by spatially resolved long-slit and integral field unit observations \citep{nesv06, nesv08, gree11, cano12, liu13a, liu13b, alex13, rupk13a, harr14, zaka14}. Gas extended over a few kpc scales and moving with velocities of $\ga 1000$ km s$^{-1}$ cannot be in dynamical equilibrium with the galaxy and cannot be confined by any realistic galactic potential. Even in the absence of spatial information, it is often assumed that in quasars forbidden line emission must be extended on scales of $\ga 1$ kpc because forbidden transitions arise in relatively low density ($n$ from 10 cm$^{-3}$ to a few$\times 10^5$ cm$^{-3}$) warm ($T\sim 10^4$~K) clouds. Therefore, very high velocity dispersions or strong line asymmetries in forbidden lines -- especially blue-shifts which arise naturally in dusty winds \citep{whit85a} -- are sufficient evidence of high-velocity extended outflows \citep{spoo09, mull13, zaka14, brus15}. 

The key epoch for studies of quasar feedback is that of peak star formation and quasar activity at redshifts $z=2-3$ \citep{boyl98, hopk08}: this is likely the epoch at which the modern-day relationships between black holes and their hosts were established, as massive galaxies grew most of their stellar mass at that time. In evolutionary models of quasar obscuration \citep{sand88, hopk06} it is the obscured quasars which are most likely to be found in the strong feedback phase. Therefore, luminous red and type 2 quasars at the peak galaxy formation epoch are promising locations to look for the most powerful quasar-driven winds. While obscured quasars remain difficult to find in large numbers at high redshifts because of their faintness in the rest-frame ultra-violet and optical, and often even at X-ray wavelengths, several samples have become recently available \citep{alex13, eise12, tsai15, ross15, brus15}. 

In this paper we present a discovery of extreme ionized gas outflows in four luminous red quasars identified by \citet{ross15}, which we will argue are a manifestation of strong quasar feedback. In Section \ref{sec:data} we describe our observations and data reduction. In Section \ref{sec:kin} we analyze the kinematics and the strengths of the forbidden emission lines. In Section \ref{sec:disc} we compare our results to those of other samples and discuss the possible role of such objects in galaxy evolution, and we conclude in Section \ref{sec:conc}. We use an $h=0.7$, $\Omega_m=0.3$, $\Omega_{\Lambda}=0.7$ cosmology. The sources are identified by their full hhmmss.ss+ddmmss.s coordinates in Table \ref{tab:1} and shortened to SDSS~Jhhmm+ddmm elsewhere. We use air wavelengths for emission-line identifications in the text, but because SDSS and VLT spectra are calibrated for vacuum wavelengths, in all calculations we use vacuum wavelengths of emission lines (e.g., [OIII]$\lambda$5007\AA\ has a laboratory wavelength of 5008.24\AA).

There is extensive nomenclature associated with accreting supermassive black holes. In this paper we refer to them collectively as active (galactic) nuclei, and to the luminous subset of them ($L_{\rm bol}\ga 10^{45}$ erg s$^{-1}$) as quasars. We refer to quasars as `type 1', `type 2' and `red' based on their salient optical-to-infrared properties. Type 1 quasars are those with blue optical colors and broad permitted emission lines with widths of several thousand km s$^{-1}$. Type 2 quasars have low continua and strong emission lines with the same kinematics in forbidden and permitted emission lines and with typical line widths of a few hundred km s$^{-1}$. Red quasars are identified by their high infrared-to-optical or red-to-blue ratios. The unification model of active nuclei established a reasonable association between the amount of obscuration and these properties, with type 2 / red / type 1 classification typically running from most to least obscured, and the challenges to this picture presented by the extremely red quasars are discussed in Section \ref{sec:type}.

\section{Observations and data reduction}
\label{sec:data}

The targets for this work are drawn from the sample of extremely red quasars by \citet{ross15}. Briefly, in a search of obscured (red and type 2) quasar candidates, \citet{ross15} selected objects classified as quasars in the Baryon Oscillation Spectroscopic Survey (BOSS; \citealt{daws13}) of the Sloan Digital Sky Survey (SDSS; \citealt{york00, eise11, alam15}) database, which also had extreme infrared-to-optical colors as measured by the SDSS and the Wide-field Infrared Survey Explorer (WISE; \citealt{wrig10}). From the parent sample of 256,000 optically confirmed SDSS spectra, \citet{ross15} selected and studied 65 sources at the tail of the infrared-to-optical ratio distribution.

\begin{figure}
\includegraphics[scale=0.45]{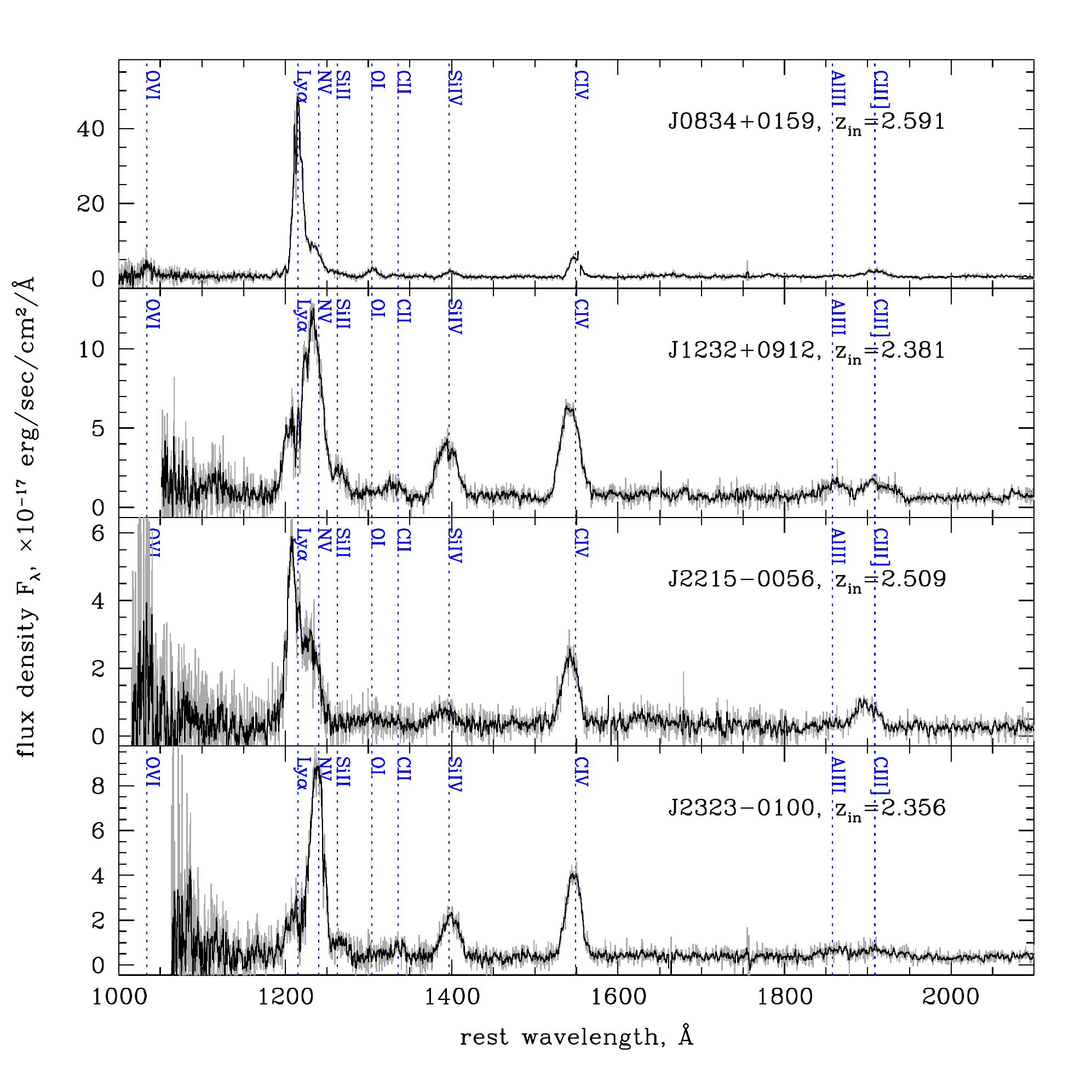}
\caption{SDSS / BOSS spectra of the four extremely red quasars with extreme forbidden-line kinematics discovered in follow-up near-infrared observations (grey: unsmoothed, black: smoothed with 5-bin boxcar filter). The spectra are corrected to the rest frame of the sources using BOSS pipeline redshifts $z_{\rm in}$, as indicated in each panel. We use these redshifts throughout the paper, keeping in mind that the true rest-frame of the sources may not be well determined by the redshifts of the emission lines which may be subject to strong outflow signatures. }
\label{pic_opt}
\end{figure}

Most of these objects are, as expected, type 2 or heavily reddened type 1 quasars, but about twelve objects defy simple explanations. These sources have rest-frame ultra-violet emission lines with the high rest equivalent widths (e.g., CIV$\lambda$1549\AA\ with REW$>100$\AA) expected in type 2 quasars where the continuum is strongly suppressed due to obscuration but lines arise in an extended region. However, the emission line velocity widths are $> 2000$ km s$^{-1}$, which is typical of type 1 quasars, and emission lines often display bizarre shapes and line ratios not commonly seen in active nuclei of any known type. Specifically, these objects show high NV$\lambda$1240\AA/Ly$\alpha$ ratios and peculiar ``stubby'' CIV$\lambda$1549\AA\ emission line profiles which lack the extended wings characteristic of lines in ordinary quasars (Figure \ref{pic_opt}). A comprehensive study of the colors and the line properties of these objects in comparison to the general quasar population will be presented by Hamann et al. (2016a, in prep.).

\begin{figure}
\includegraphics[scale=0.45]{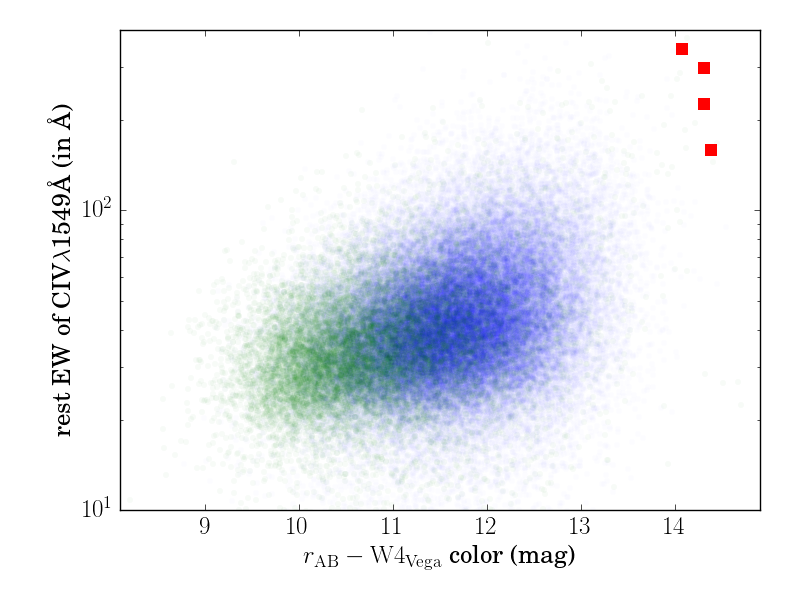}
\caption{Optical-to-infrared colors and rest equivalent widths of CIV in the four objects presented in this paper (red squares) compared to those of DR12 BOSS quasars at $2<z<3$ (\citealt{pari14}, P\^aris et al. in prep.). Green points show $\sim 18,000$ objects that are detected in W4, whereas blue points correspond to $\sim 147,000$ W4 upper limits, and therefore their $r-{\rm W4}$ colors are constrained to be smaller (bluer) than shown. Our targets stand out both in terms of their very red optical-to-infrared colors and their high REW of CIV$\lambda$1549\AA.}
\label{pic_ref}
\end{figure}

We selected four such targets with infrared-to-optical ratios $r_{\rm AB}-{\rm W4}_{\rm Vega}>14$ mag and with rest equivalent widths of CIV$\lambda$1549\AA\ in excess of 100\AA\ for observations (PI: I. P\^aris) with the European Southern Observatory's Very Large Telescope (VLT) using XShooter, a medium resolution spectrograph allowing simultaneous observations over the wavelength range from 0.3 to 2.48 \micron\ \citep{vern11}. In our upcoming analysis of the full VLT dataset (Hamann et al. in prep. 2016b) we will discuss the optical part of the VLT spectra and comparison of optical properties of quasars of different infrared-to-optical colors. In this paper we focus on the spectacular forbidden [OIII]$\lambda\lambda$4959,5007\AA\AA\ emission lines discovered in the four red quasars using the near-infrared part of the VLT spectra. The BOSS spectra of the four targets discussed in this paper are shown in Figure \ref{pic_opt}, and the comparison of our targets with other BOSS quasars in colors and CIV equivalent widths is shown in Figure \ref{pic_ref}. None of the objects are detected in the FIRST survey at 1 mJy level \citep{beck95}, implying that their radio luminosities are below $\nu L_{\nu}$[1.4GHz]$\sim 5\times 10^{41}$ erg s$^{-1}$.

We acquired near-infrared spectra with slit width of 0.9\arcsec, resulting in a spectral resolution of $R=5100$ \citep{vern11}. Data were acquired in service mode between 2nd April 2014 and 26th July 2014. Our program was designed to have a homogeneous signal-to-noise across the sample based on the known optical fluxes which range between 21.1 and 22.3 mag in the $r$-band (Table \ref{tab:1}). The total exposure time for brighter targets (\oviii\ and \oxii) was 2400s, and it was 3600s for the fainter ones (\oxxii\ and \oxxiii). The sources are covered by the UKIDSS near-infrared survey \citep{lawr07} but each is only detected in one or two bands with large photometric uncertainties, with $Y$-band Vega magnitudes fainter than 19th and $K$-band Vega magnitudes fainter than 17th.

We reduced the data using a custom pipeline developed by George Becker and described in Lopez et al. (in prep.). It is based on the techniques of \citet{kels03} with all the steps computed on the unrectified two-dimensional frames. Each near-infrared frame was dark subtracted and flat-fielded. Sky emission in each order was modeled with a b-spline and was subtracted. Even in the near-infrared arm, sky emission could be well modeled in each exposure, without subtracting a nodded frame. This procedure has the advantage of avoiding a $\sqrt{2}$ penalty in the background noise. The reddest order (2.27$-$2.48 \micron) had to be nod-subtracted. The resulting signal-to-noise in the last order is lower than in the rest of the spectrum and the relative flux normalization between that order and the rest of the spectrum is uncertain. The residual sky emission was modeled with a b-spline. Median per-pixel uncertainties resulting from the pipeline are in good agreement (within 30\%) with the standard deviation of data around a linear continuum in a line-free region.

The counts in the two-dimensional frames were flux-calibrated using response curves generated from the observations of spectro-photometric standard stars. A single one-dimensional spectrum was then extracted simultaneously across all the orders and all exposures of a single object, avoiding multiple rebinnings and minimizing the error correlation between adjacent pixels. The one-dimensional spectra were binned to have fixed-velocity pixels (19 km s$^{-1}$). The wavelengths are on the vacuum heliocentric system. 

Telluric corrections were performed using {\sc Molecfit} \citep{smet15, kaus15}. Because the targets were specifically selected in redshift to avoid an overlap between the major emission lines and the strong telluric absorption features, the effect of telluric corrections is small for most measurements. The only noticeable effect is in \oviii, where telluric corrections lead to a 60\% increase in the measured [OIII]$\lambda$5007\AA\ flux.

\section{Analysis of spectra and spectral energy distributions}
\label{sec:kin}

\subsection{[OIII] kinematics}
\label{sec:oiii}

\begin{figure*}
\centering
\includegraphics[scale=0.7]{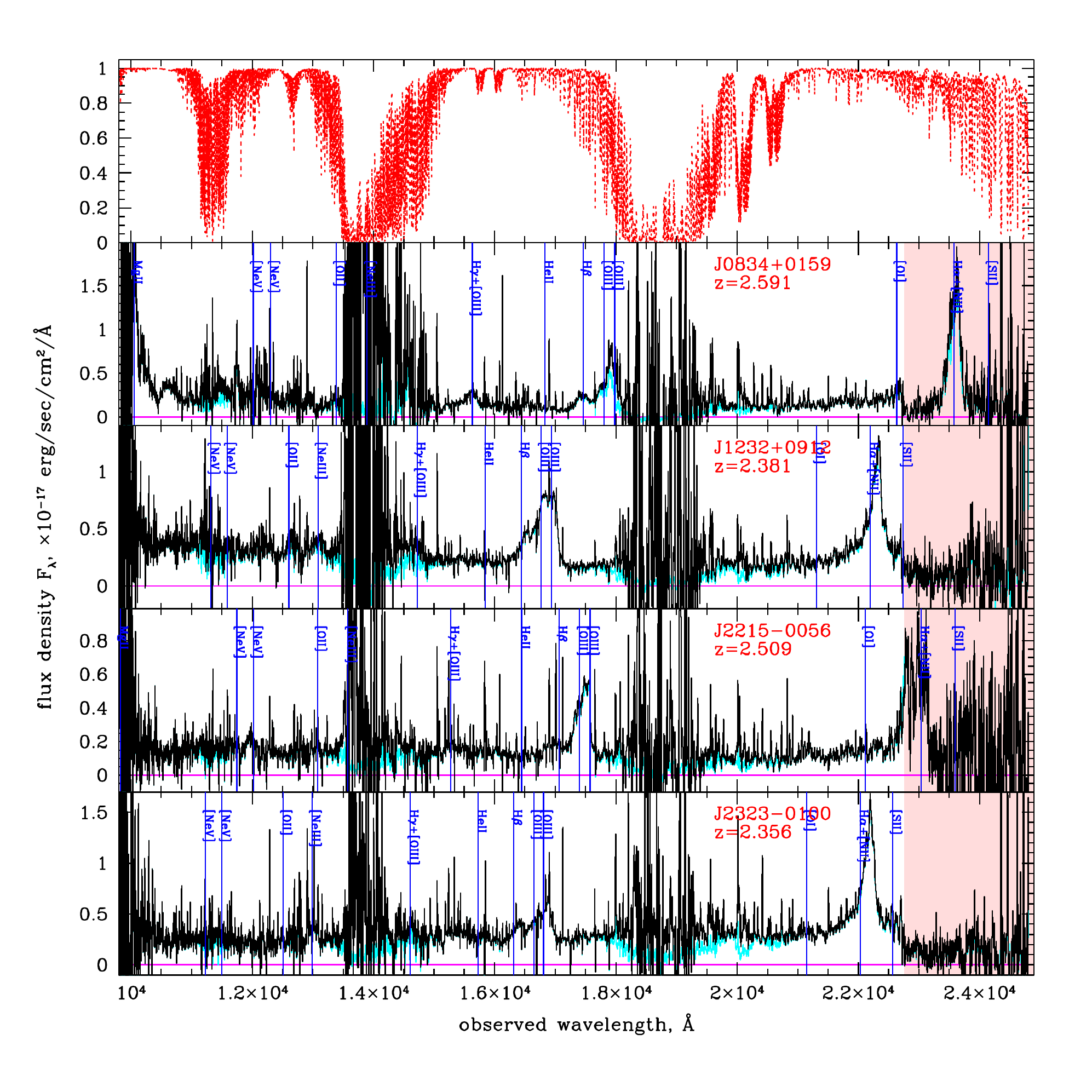}
\caption{Top panel: atmospheric transmission curve for SDSSJ0834+0159. Bottom four panels: VLT spectra of the four extremely red quasars in this paper. Cyan histogram shows extracted one-dimensional spectra smoothed with a 10-pixel boxcar, and black shows the telluric-corrected, 10-pixel-smoothed spectrum (atmospheric corrections are appreciable where the two are distinguishable). The spectral regions of poor opacity correspond to the gaps between $J$, $H$ and $K$ atmospheric bands, where the reconstructed spectrum has high noise. Blue vertical lines mark the expected locations of the major rest-frame-optical emission lines if the BOSS pipeline redshift is adopted at face value. The pink shaded region indicates the last order ($\lambda > 22753$\AA) where the flux calibration relative to the rest of the spectrum is uncertain. }
\label{pic_spec}
\end{figure*}

Figure \ref{pic_spec} shows our VLT spectra of four extremely red quasars. The extreme kinematics of the [OIII]$\lambda\lambda$4959,5007\AA\AA\ emission lines are immediately apparent. Not only are the two lines of the [OIII] doublet blended together (which does occur occasionally in the most kinematically extreme quasars at $z\la 1$, \citealt{zaka14}), the doublet is further blended with H$\beta$, which indicates that line-of-sight velocities of order several thousand km s$^{-1}$ must be present. 

In this section, we quantify the kinematics of the [OIII] emission. Ideally, the best way to measure velocities would be relative to the rest-frame of the host galaxy as measured by stellar absorption lines, but we see no such lines in our spectra -- the optical and near-infrared spectra are likely dominated by the quasar. Therefore, as an initial approximation we adopt the nominal BOSS pipeline redshifts $z_{\rm in}$ (listed in Table \ref{tab:1}) which are obtained by cross-correlating the optical (rest-frame ultra-violet) spectra against pre-defined templates (\citealt{bolt12}; in our case, against a normal blue type 1 quasar template), but with an understanding that they are not necessarily accurate. We then measure all kinematics relative to that frame and discuss possible avenues for determining the absolute host galaxy frame later in this section. 

With such extremely high velocities in the forbidden emission lines, the traditional use of line kinematics for quasar type determination (``narrow line'' type 2 vs ``broad line'' type 1) is meaningless (Section \ref{sec:type}). The more physically motivated question is whether [OIII] and H$\beta$ have consistent kinematic structure, which might indicate that they are emitted in the same physical region. Therefore, for every object we explore two classes of emission line fits: (1) `kinematically tied' fitting functions which assume the same kinematics for H$\beta$ and each of the [OIII] lines, and (2) `kinematically distinct', or `untied', fitting functions in which H$\beta$ is allowed to have different kinematic structure from the components of the [OIII] doublet. The components of the [OIII] doublet are always assumed to have the same kinematics and the $F_{\lambda}$ amplitude ratio [OIII]$\lambda$4959\AA/[OIII]$\lambda$5007\AA\ of 0.337. 

To determine the underlying continuum, we calculate the median flux in 20\AA-wide windows centered at rest-frame 4750\AA\ and 5090\AA, then subtract a linear interpolation between these two points to obtain the [OIII]+H$\beta$ emission-line blend. Using one-Gaussian fits, we find that fits with kinematically distinct [OIII] and H$\beta$ are preferred for \oviii, \oxii, and \oxxiii, whereas for \oxxii\ there is no appreciable statistical or visual difference between tied and untied fits. Specifically, in \oviii, \oxii, and \oxxii\ the centroids of [OIII] (as measured from the single-Gaussian fits) are offset from centroids of H$\beta$ by -1009, -1907 and -1545 km s$^{-1}$, respectively (in \oxxii, the offset is nominally measured at 128 km s$^{-1}$, which is consistent with 0 given the large line widths of all features). It is also clear from these crude fits that the kinematics of [OIII] are rather extreme, with velocity dispersions 1580, 2002, 1272 and 1168 km s$^{-1}$ for the four objects. Single-Gaussian fits do not capture well the shape of the lines, and in relatively low signal-to-noise spectra such fits lead to an underestimate of the actual velocity width \citep{zaka14}.

\begin{figure*}
\includegraphics[scale=0.9, clip=true, trim=0cm 6cm 0cm 0cm]{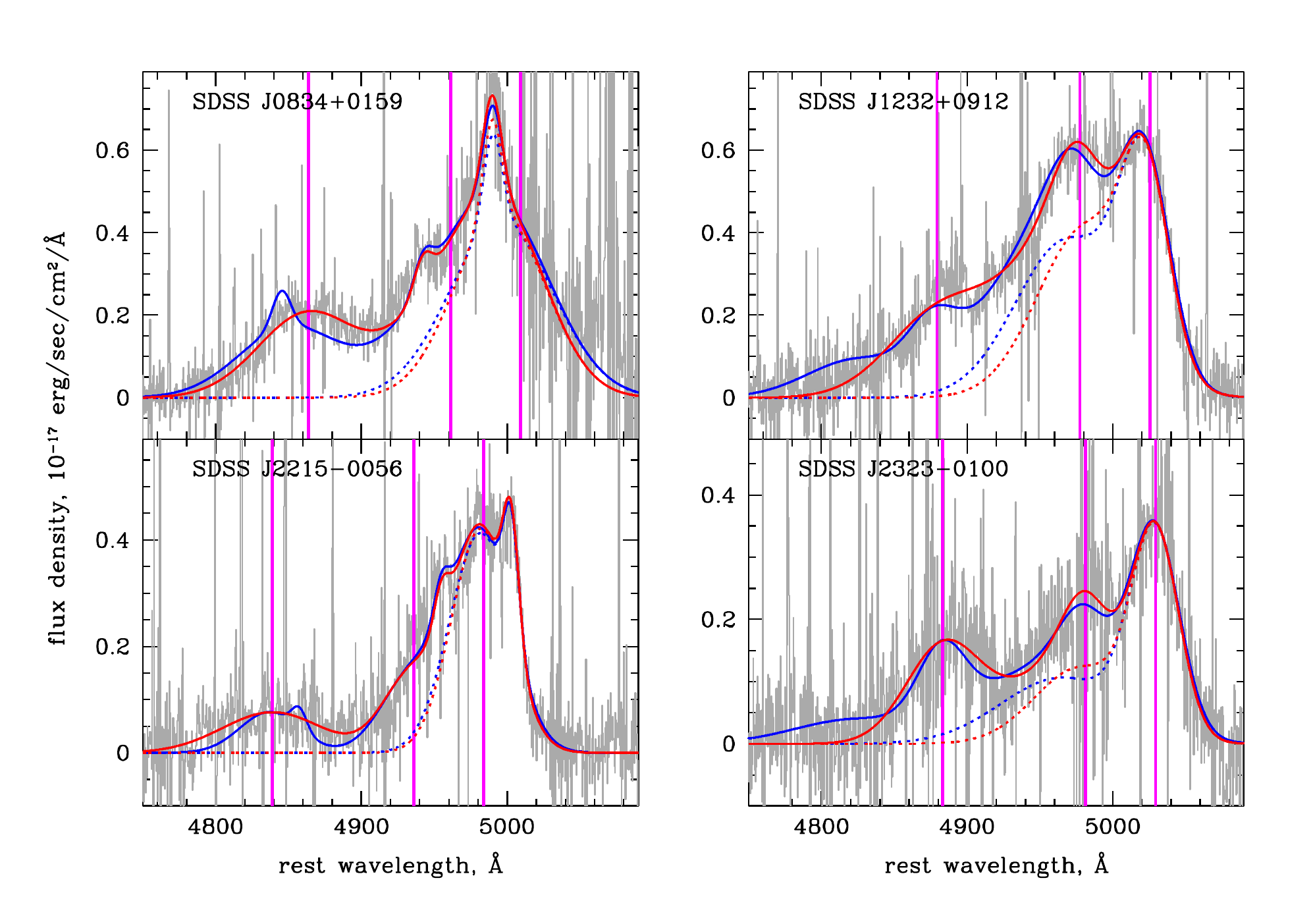}
\caption{Fits to the H$\beta$+[OIII] blend in the four extremely red quasars observed with VLT. A linear continuum anchored at 4750\AA\ and 5090\AA\ is subtracted before fitting. The blue solid line shows the best two-Gaussian, kinematically tied fit to the entire complex, whereas the red line assumes two Gaussian components for [OIII] and a single kinematically independent Gaussian for H$\beta$. The dotted lines display just the [OIII]$\lambda$5007\AA\ profile for both fits. The solid magenta vertical lines mark the locations of H$\beta$, [OIII]$\lambda$4959\AA\ and [OIII]$\lambda$5007\AA\ in the frame associated with the centroid of H$\beta$ in the kinematically untied fit. The red line is our preferred fit for \oviii, \oxii\ and \oxxiii, and the blue line is the preferred fit for \oxxii\ (in the latter case there is no statistical difference between the two fits, thus the one with the lower number of parameters is preferred).}
\label{pic_oiii}
\end{figure*}

All fits are improved by adding a second Gaussian component (average reduced $\chi^2$ improvement is 20\%), and these fits are presented in Figure \ref{pic_oiii}. The kinematically tied 2-Gaussian fits (shown in blue) assume that [OIII]$\lambda$4959\AA, [OIII]$\lambda$5007\AA\ and H$\beta$ have the same kinematic structure -- i.e., all three lines have the same velocity offsets and velocity dispersions for each of the two Gaussian components and the same amplitude ratio of the two Gaussian components. The kinematically untied 2-Gaussian fits assume the same 2-Gaussian kinematic structure for [OIII]$\lambda$4959\AA\ and [OIII]$\lambda$5007\AA, but allow for a separate unconstrained single Gaussian for H$\beta$. Because of the relatively low signal-to-noise ratio of the H$\beta$ detection alone, there is no statistical need for adding a second Gaussian component to H$\beta$ in the kinematically untied fits. 

The large apparent ratio of peaks of [OIII]$\lambda$4959\AA\ and [OIII]$\lambda$5007\AA\ -- close to 1:1 in \oxii, compared to the theoretical ratio of 0.337 -- implies that a strong broad blue-shifted component is required to fit the [OIII] doublet, which boosts the flux near the peak of [OIII]$\lambda$4959\AA. In \oviii, \oxii, and \oxxiii\ we find that kinematically untied fits produce a slightly better fit to the data; the main reason is that there is no evidence in the data for the broad blue component to the H$\beta$ line of the kind that is clearly required to fit the [OIII] doublet. This is especially clear in \oviii, where the kinematically tied fit fails for the entire H$\beta$ profile -- both its blue and its red side. In \oxxii\ we find no statistically significant difference between kinematically tied and kinematically untied fits, so we prefer the kinematically tied fit for this object as the more simple model.  

We do not assign any particular physical meaning to the parameters of the individual Gaussian components within the two-Gaussian fits for [OIII]. Rather, we use the fits to compute the non-parametric measures of the [OIII]$\lambda$5007\AA\ profile, following \citet{zaka14} and many other authors (e.g., \citealt{whit85a, veil91b}). For any functional fit to the [OIII]$\lambda$5007\AA\ profile, we can calculate the velocity width comprising 90\% of the flux $w_{90}$ (and similarly $w_{80}$) by rejecting the most extreme 5\% blue-shifted and red-shifted parts of the profile. For a Gaussian profile, both are related to the velocity dispersion, with $w_{80}$ being quite close to the commonly used full width at half maximum ($w_{80}=2.563\sigma_v=1.088$FWHM; $w_{90}=3.290\sigma_v$). We list these measurements, together with full width at half maximum (FWHM) and full width at quarter maximum (FWQM) in Table \ref{tab:1}. 

\begin{figure}
\includegraphics[scale=0.45, clip=true, trim=0cm 6cm 0cm 0cm]{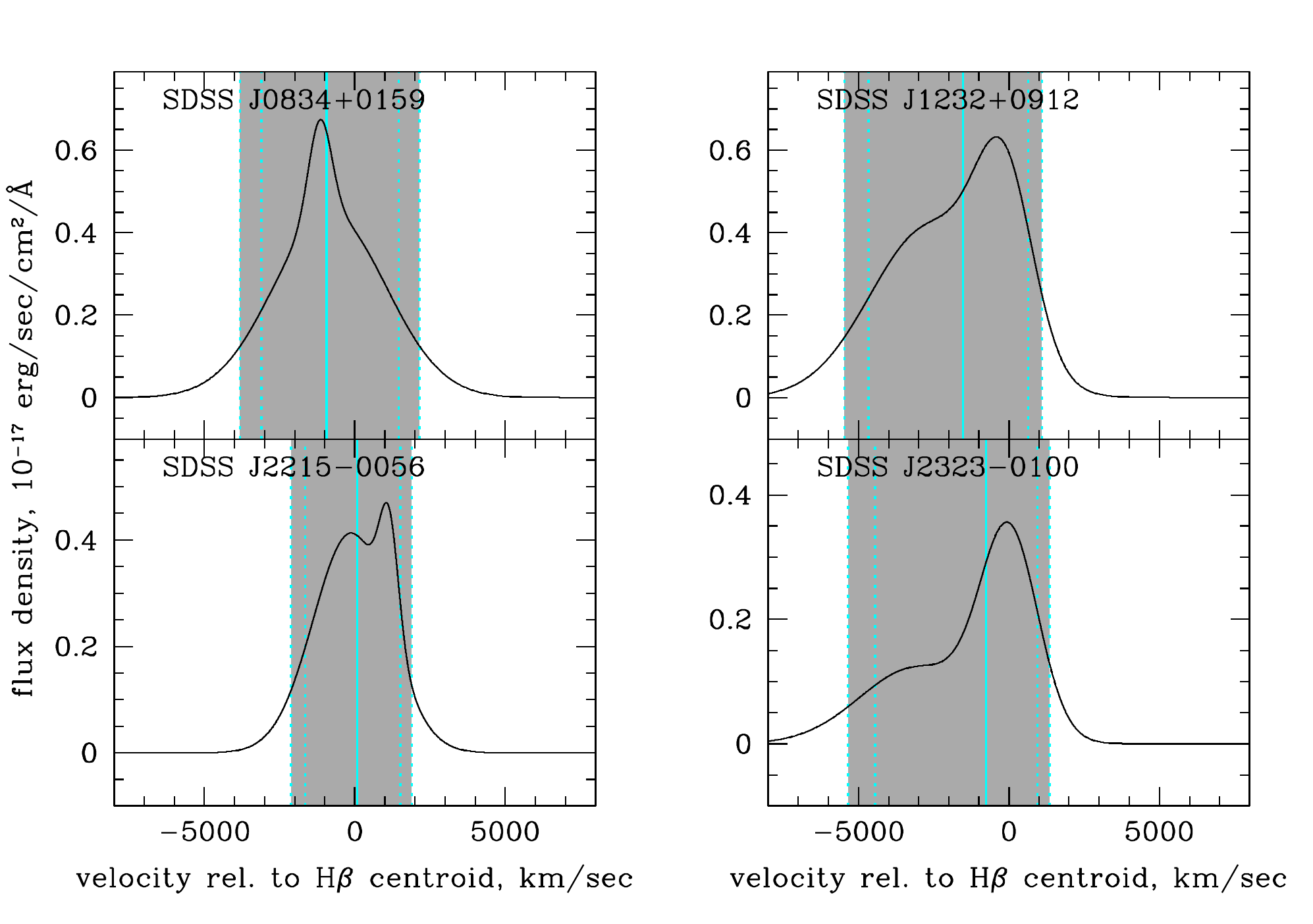}
\caption{The best-fit profiles of [OIII]$\lambda$5007\AA\ relative to the centroid of H$\beta$, shown in velocity space. Solid grey marks the part of the profile containing 90\% of the line power ($w_{90}$), whereas vertical cyan lines mark $v_{05}$, $v_{10}$, $v_{90}$, $v_{95}$ (dotted) and the median velocity of the profile $v_{50}$ (solid).}
\label{pic_nonparam}
\end{figure}

The method is illustrated in Figure \ref{pic_nonparam}, where we display the best-fit [OIII]$\lambda$5007\AA\ profiles in velocity space relative to the centroid of the H$\beta$ determined from the kinematically untied fits. As measured relative to the frame of the H$\beta$ centroid, median velocities $v_{50}$ (in the order of increasing object right ascension) are $-937$, $-1520$, $+79$, and $-769$ km s$^{-1}$, i.e., [OIII] is strongly blueshifted relative to the measured H$\beta$ centroid in all but one case (\oxxii). The velocity widths $w_{80}$ range between 3600 and 5500 km s$^{-1}$,  FWHM between 2600 and 5000 km $^{-1}$ and $w_{90}$ between 4000 and 6700 km s$^{-1}$. Such [OIII] widths are completely outside the range found in type 1 quasars \citep{stei13, shen16} and type 2 quasars at low redshifts \citep{zaka14}, as further discussed in Section \ref{sec:type}. The peak signal-to-noise within the [OIII] doublet is $\sim 10$ in our sources. If our assumptions about H$\beta$ kinematics are correct, simulations conducted by \citet{zaka14} suggest that $w_{80}$ and $w_{90}$ are reliable for these values and accurate within $\la 15\%$ and that, if anything, these values would be underestimated in our sources because we may have missed another weak broad component.  

[OIII] can be artificially broadened by FeII contamination in type 1 quasars. While in type 2 quasars FeII is not seen, we allow for the possibility that we see at least some nuclear continuum in our objects (whether scattered or directly penetrating through patchy obscuration; Section \ref{sec:sed}). To estimate the effect of FeII on our kinematic measurements, we use the Fe template from \citet{boro92} which we convolve with a Gaussian function with the velocity dispersion of H$\beta$. Because the convolved templates show characteristic Fe blends centered at $\sim 4600$\AA\ and $\sim 5300$\AA, by looking at the maximal possible amplitudes of these features in the spectra we can subtract the maximal plausible Fe template from the data and remeasure the kinematics of the [OIII]+H$\beta$ blend. We find that the line widths of [OIII] decrease by 10\% in \oxxiii, 8\% in \oxxii\ and 2\% in \oxii, and increases by 8\% in \oviii. We conclude that FeII does not severely affect the measured line kinematics. Furthermore, this allows us to estimate that the systematic uncertainties in our fits (e.g., due to continuum placement) are $\la 10\%$ for the kinematic measures presented in Table \ref{tab:1}. 

The FWHM of H$\beta$ ranges between 3700 and 5000 km s$^{-1}$. This is in the normal range for type 1 active nuclei and quasars \citep{hao05a, stei13}. Because of the specifics of our fitting routine, an extremely broad redshifted H$\beta$ component would be fitted out as the blue-shifted component of [OIII]. Because we see no evidence for such component in H$\alpha$ (Section \ref{sec:nii}), we find this scenario unlikely and we take the H$\beta$ fit parameters at face value. 

\subsection{[NII] and H$\alpha$}
\label{sec:nii}

\begin{figure*}
\includegraphics[scale=0.9, clip=true, trim=0cm 6cm 0cm 0cm]{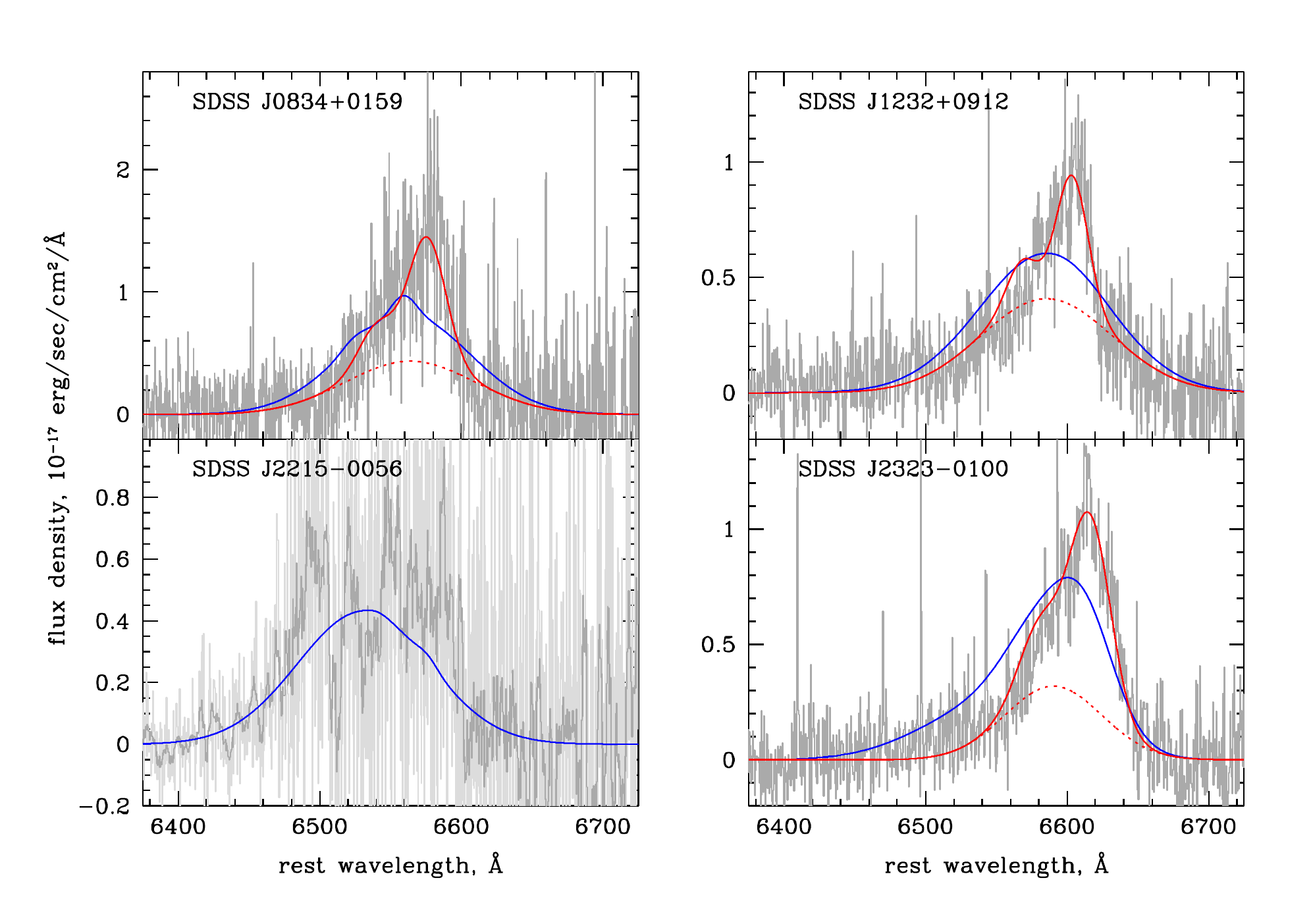}
\caption{Fits to the H$\alpha$+[NII] blends in the frame of the BOSS pipeline redshift, shown with linear continuum subtracted. We carry out two fits, both with H$\alpha$ constrained to have the same shape as H$\beta$. In the first (blue), we tie the [NII] lines to the [OIII] profile, while in the second (red) the kinematics of [NII] are fit by a single Gaussian. Dotted lines show the contribution of H$\alpha$ in the second fit. In \oxxii, adding [NII] lines to the fit does not improve the $\chi^2$ and we consider them undetected. In this object, light grey shows unbinned spectrum and dark grey the 10-pixel-binned one.}
\label{pic_nii}
\end{figure*}

[NII]$\lambda\lambda$6548,6583\AA\ and H$\alpha$ are closer together in wavelength than [OIII] and H$\beta$, so disentangling their kinematics is even more complex. Therefore, we first attempt to fit this line blend with the best-fit kinematics of the [OIII]+H$\beta$ blend. We assume that H$\alpha$ has the same kinematics as H$\beta$ (one-Gaussian fit) and that H$\alpha$ flux is $\ge 2.85\times$ H$\beta$ as required by the Case B recombination with allowance for reddening. Further, we assume that both [NII] lines have the same kinematics as [OIII] and that the peak $F_{\lambda}$ ratio of [NII]$\lambda$6548\AA\ to [NII]$\lambda$6583\AA\ is 0.340. With these shape constraints, there are only two fitting parameters -- the amplitudes of H$\alpha$ and [NII]$\lambda$6583\AA. In three cases where the noise is not prohibitively large, fits with these constraints fail in that they produce much more blue-shifted emission than what is observed (Figure \ref{pic_nii}). From this we conclude that the [NII] emission must be not nearly as blue-shifted and broad as [OIII] and thus likely arises from a different spatial region in these quasars. 

In \oviii, \oxii\ and \oxxiii, we successfully fit the [NII]+H$\alpha$ blend with a model that uses the same constraints on H$\alpha$ as above (one-Gaussian kinematics fixed to that of H$\beta$, with Case B constraint on the minimal H$\alpha$ flux) but allows for kinematics of [NII] independent from those of [OIII] (Table \ref{tab:1}). One Gaussian component for [NII] is sufficient to produce adequate fits. In all three objects, the final H$\alpha$/H$\beta$ ratio is consistent with Case B values, even though a larger ratio is allowed by the fit, and in fact without the lower bound the preferred H$\alpha$/H$\beta$ ratios are smaller by as much as 50\% than the Case B values (the most discrepant object is \oviii, where unfortunately the absolute flux calibration of the spectral region covering H$\alpha$ is the most uncertain). One possible explanation is that in fact the true H$\alpha$/H$\beta$ ratios are smaller than Case B, which is difficult \citep{gask84}, but not impossible \citep{kori04}. Another possibility is that the kinematics of H$\alpha$ are different from the kinematics of H$\beta$. A narrower H$\alpha$ which is slightly redshifted relative to H$\beta$ might improve the fits, but the blending of H$\alpha$ with [NII] prevents us from testing such models. 

In \oxxii, we successfully measure an H$\alpha$/H$\beta$ ratio between 7.6 and 11.4, with the range being due to the low signal-to-noise ratio measurement of the H$\beta$ line and arising from the difference in the derived fluxes in the kinematically tied and kinematically untied fits which are both acceptable. Using the \citet{wein01} Small Magellanic Cloud extinction curve and assuming an intrinsic H$\alpha$/H$\beta$ ratio of 2.85, we derive a corresponding extinction of the Balmer-emitting region of $A_V=2.6-3.7$ mag. H$\alpha$ is marginally better reproduced by the kinematically untied fits to the H$\beta$+[OIII] complex; when H$\alpha$ is fit with just the [OIII] profile, the blueshifted side of the H$\alpha$ emission is not well reproduced. From this we might conclude that the Balmer emission originates in the broad-line region which is kinematically distinct from the [OIII]-emitting region and that therefore the derived extinction values apply to the broad-line region in this object. However, H$\beta$ is very weak in this source and H$\alpha$ is in the part o the spectrum with a poor signal-to-noise ratio, so these conclusions are only tentative.

\begin{figure}
\includegraphics[scale=0.7, clip=true, trim=0cm 10cm 10cm 0cm]{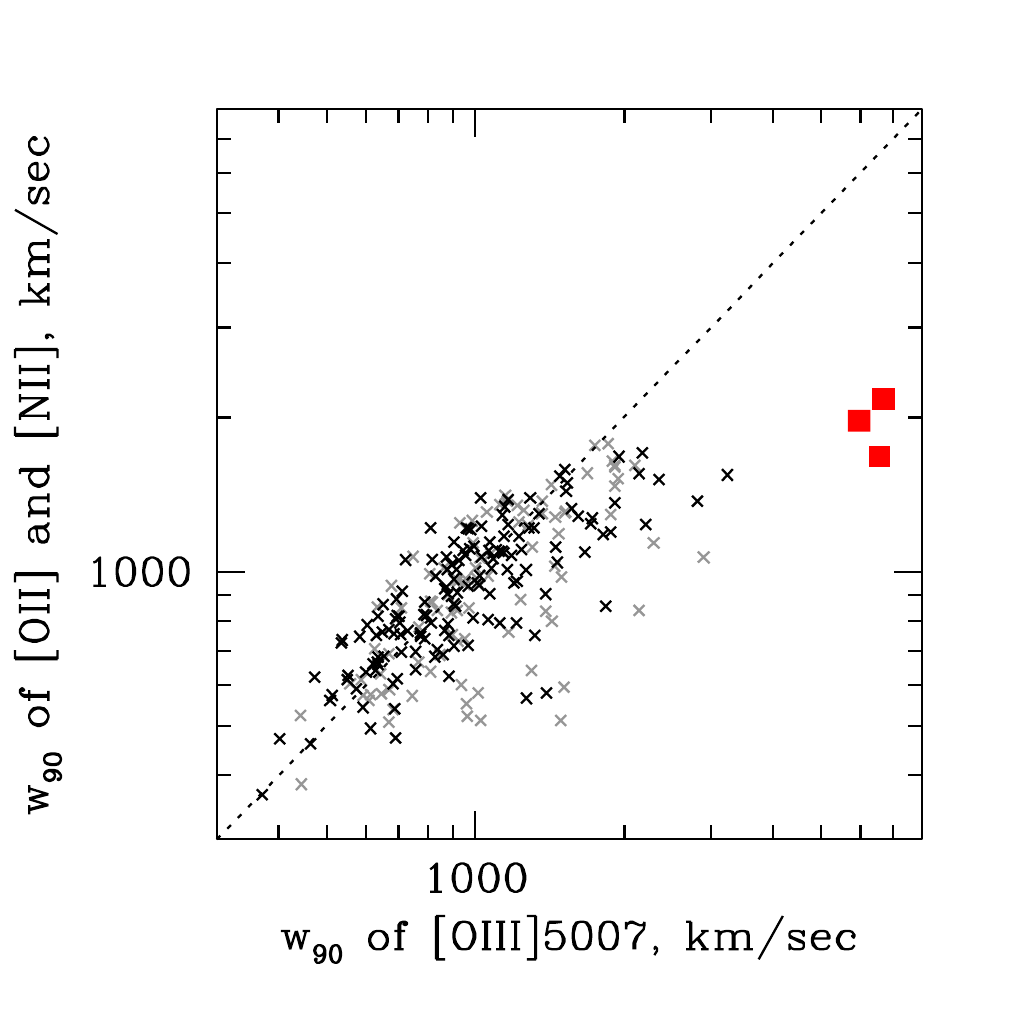}
\caption{In low-redshift type 2 quasars, the [OII]$\lambda\lambda$3726,3729\AA\AA\ doublet tends to be more kinematically quiescent than the [OIII] emission line (black crosses for high signal-to-noise [OII] detections, grey for lower signal-to-noise ratio, data from \citealt{zaka14}). This trend is also seen in the [NII] emission lines in our extremely red quasars (red squares). The measured [OII] widths have been corrected for doublet splitting as described by \citet{zaka14}. The dotted line shows the 1:1 relation.}
\label{pic_width}
\end{figure}

With the exception of \oxxii, where we consider [NII] to be non-detected, the best-fit [NII] lines are significantly narrower than the best-fit [OIII]. This is shown in Figure \ref{pic_width}, where we use type 2 quasars at $z\la 1$ with [OII]$\lambda$3726,3729\AA\AA\ kinematics for comparison as the ionization potential of [NII] (14.5 eV) is similar to that of [OII] (13.6 eV). This observation is in line with studies demonstrating that lower ionization lines have more quiescent kinematics than do higher ionization ones \citep{whit85c, dero86, veil91c}. 

The velocity centroids of [NII] (when fitted independently of [OIII]) are offset by $-100$ to $300$ km s$^{-1}$ from the centroids of H$\alpha$ and H$\beta$ (the measurement uncertainty in the centroid velocity is $\la 100$ km s$^{-1}$), whereas in the same three objects the $v_{50}$ centroids of [OIII] relative to the Balmer lines are between $-800$ and $-1500$ km s$^{-1}$. In the absence of the detection of the stellar absorption features, the centroids of the lowest-ionization forbidden emission lines are perhaps the best indicator of the host redshift. It is encouraging that the [NII] centroids are quite close to the Balmer centroids in the three objects. Although Balmer lines which originate close to the black hole may be affected by outflows (e.g., \citealt{floh12}), kinematics of the disk around the black hole (e.g., \citealt{erac94, stra03}) and gravitational redshifts (e.g., \citealt{trem14}), in practice they appear to be good proxies for the host redshifts. From the values tabulated in Table \ref{tab:1}, the overall host redshifts can be calculated as $z_{\rm true}=z_{\rm in}\left(1+v_{\rm H\beta}/c\right)$.

\subsection{Extreme [OIII]$\lambda$4363\AA\ in \oviii}
\label{sec:4363}

The [OIII]$\lambda$4363\AA\ transition originates from a higher energy state than do [OIII]$\lambda\lambda$4959,5007\AA\AA. As a result, the line ratio $R_{\rm [OIII]}\equiv(F_{4959}+F_{5007})/F_{4363}$ is a frequently used temperature diagnostic \citep{oste89}, normally ranging between 250 and 40 for temperatures between 10,000 K and 20,000 K and observed to be $\sim 90$ in low-redshift type 2 quasars \citep{zaka14}. 

By comparison to this typical value, \oviii\ displays extraordinarily strong [OIII]$\lambda$4363\AA\ emission (Figure \ref{pic_spec}). Given the extreme kinematics of the emission lines in this source, [OIII]$\lambda$4363\AA\ is blended with H$\gamma$ at 4341\AA: the nominal velocity separation between the two lines is $\sim 1500$ km s$^{-1}$, which is not sufficient to separate them out in this case. A precise measurement of $R_{\rm [OIII]}$ is therefore not possible, but we are able to recover robust lower and upper bounds as discussed below. 

We try several approaches to deblending. In all fits we assume that the kinematics of H$\gamma$ are the same as the kinematics of H$\beta$. We either fix H$\gamma$ to its Case B value (H$\gamma$/H$\beta$=0.46) or allow for a moderate extinction $A_{\rm V}\le 2$ mag (even though no extinction is preferred by the H$\alpha$ fit, the flux calibration of the longest wavelength order is uncertain enough that we do not fully rely on this measurement). As for [OIII]$\lambda$4363\AA, we either fix its kinematics to those of [OIII]$\lambda$5007\AA\ or fit it with a single Gaussian not kinematically constrained to any other lines. 

As the total flux of the blend is fixed by the data, these assumptions provide us with a lower and an upper bound on the H$\gamma$ flux and therefore on the upper and a lower bound on the [OIII]$\lambda$4363\AA\ flux. Our resulting bounds on $R_{\rm [OIII]}$ are $13\le R_{\rm [OIII]}\le 39$. What makes these values unusual is that for densities thought to be associated with the forbidden-line region of active galaxies, $n_e\ll 10^5$ cm$^{-3}$, the derived range of $R_{\rm [OIII]}$ translates into the range of electron temperatures of $20,000\le T_e\le 64,000$ K, which is too high for a photo-ionized region in thermal balance \citep{oste89, bask05}. Another unusual detail is that there is some evidence for a difference in the kinematics of [OIII]$\lambda$4363\AA\ and those of [OIII]$\lambda$5007\AA. Specifically, fits with [OIII]$\lambda$4363\AA\ kinematics fixed to those of [OIII]$\lambda$5007\AA\ produce blue excess by comparison to the observed profile, and somewhat better fits are achieved with a less blueshifted ($-750$ km s$^{-1}$) and narrower ($\sigma_v\la 800$ km s$^{-1}$) [OIII]$\lambda$4363\AA. Because of the strong blending, we consider this evidence tentative. 

In light of the extreme [OIII]$\lambda$5007\AA\ kinematics in our sources, one possible explanation for the relatively low $R_{\rm [OIII]}$ in \oviii\ is contribution from fast shocks to the emission line ionization \citep{dopi95}. It is then particularly intriguing that \oviii\ displays a possible [OI]$\lambda$6300\AA\ emission line (Figure \ref{pic_spec}), another feature indicative of fast shocks. Alternatively, the low values of $R_{\rm [OIII]}$ may arise in regions of high electron density $n_e\ga 10^5$ cm$^{-3}$ at modest temperatures $<20,000$K \citep{bask05}. \citet{ster14} calculate line fluxes in models of radiation-pressure confined clouds, finding that $R_{\rm [OIII]}=20$ would be achieved at densities of $10^4$ cm$^{-3}$ at a distance of 120 pc$\times (L_{\rm ionizing}/10^{45}\mbox{ erg s}^{-1})^{1/2}$ from the nucleus. If this is the case, then [OIII]$\lambda$5007\AA\ might primarily arise in less dense, more easily accelerated gas, whereas [OIII]$\lambda$4363\AA\ arises in pockets of denser gas with higher inertia which are not as easily entrained by the wind, accounting for the difference in the observed kinematics.

With our current sample size and quality of the data we are unable to confidently detect [OIII]$\lambda$4363\AA\ in other sources, though there may be hints for this line in \oxii\ and \oxxiii. Unfortunately, in both cases the presence or absence of the feature is very sensitive to the assumed correction for the atmospheric transparency.

\subsection{Spectral energy distributions}
\label{sec:sed}

\begin{figure}
\includegraphics[scale=0.45]{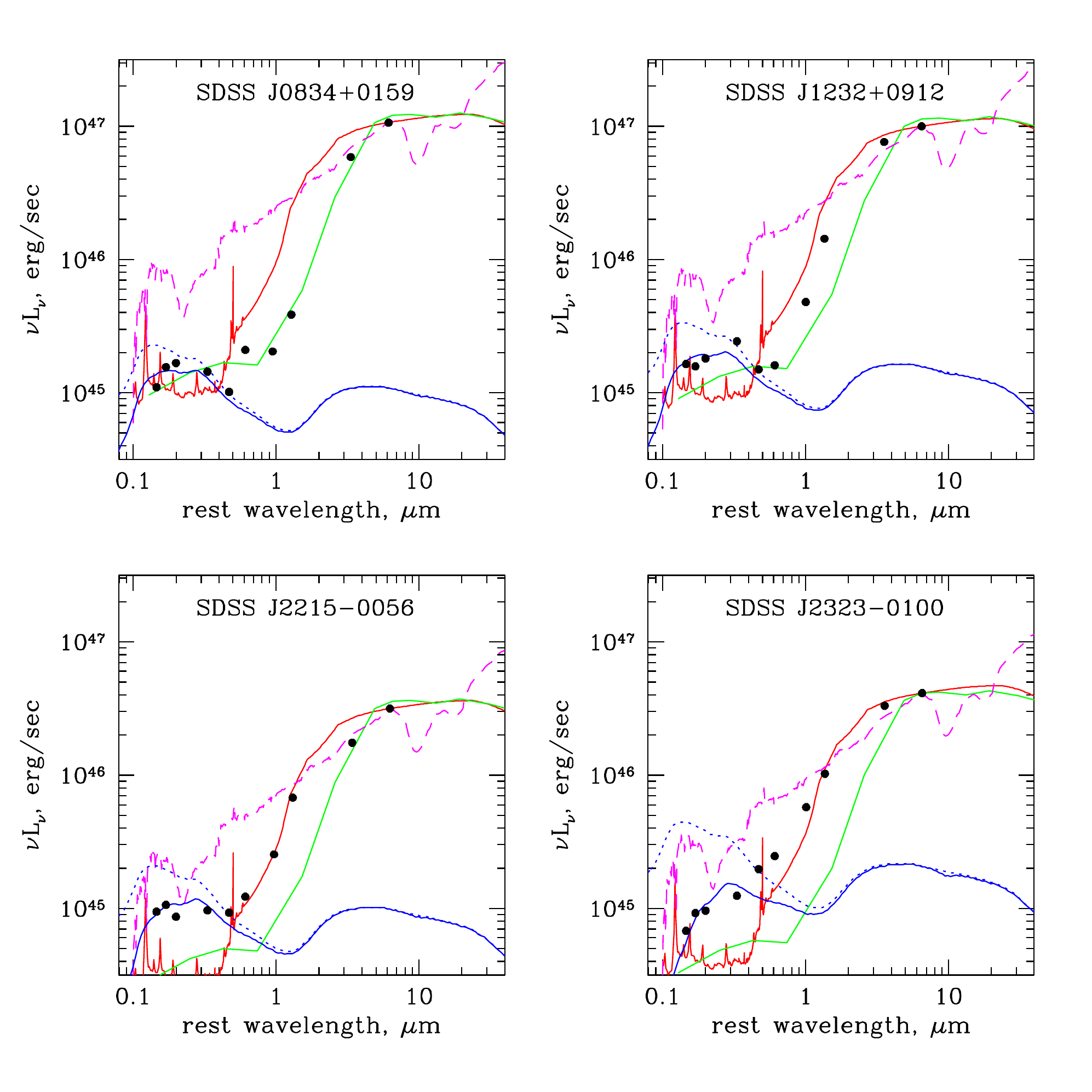}
\caption{Continuum spectral energy distributions of our targets (black data points) assembled from BOSS spectra, VLT spectra and WISE forced photometry. For comparison, red lines show ``torus'' templates from \citet{poll07} and green lines show HotDOG spectral energy distributions from \citet{tsai15}, both normalized to the longest wavelength point (W4). Optical and ultraviolet colors are inconsistent with those of type 1 quasars, with reddening by $A_V=0.6-2.5$ mag somewhat improving the agreement (solid blue lines: type 1 quasar template from \citealt{rich06} affected by dust extinction from \citealt{wein01}; dotted blue lines: templates with no extinction). The optical and ultraviolet fits are produced by quasars with bolometric luminosities $L_{\rm bol}\la 10^{46}$ erg s$^{-1}$, inadequate by at least an order of magnitude to explain the observed infrared emission. The dashed magenta template (Mrk231 from \citealt{poll07}) is a typical power-law active galactic nucleus template inconsistent with the steep mid-infrared rise of the SEDs we see in our objects. }
\label{pic_sed}
\end{figure}

For every object, we select 20\AA-wide emission-line-free windows centered at rest-frame 1460, 1700, 2000, 3300, 4700, and 6200\AA. Using BOSS and VLT spectra we extract median fluxes at these wavelengths, which we consider to be pure continuum fluxes. We supplement those with WISE forced photometry \citep{lang14} available in bands W1 (3.6\micron), W2 (4.5\micron), W3 (12\micron) and W4 (22\micron), and we show the continuum spectral energy distributions (SEDs) from optical to infrared wavelengths in Figure \ref{pic_sed}. 

Our targets display high infrared-to-optical ratios (which reflects the initial selection criteria of the parent sample) with a characteristic steep rise of the SED from 1 to 10 \micron. Of all quasar templates in the SWIRE library \citep{poll07} the only one that produces a similarly steep rise in the mid-infrared is the ``torus'' template derived from a high-redshift Compton-thick quasar \citep{poll06}, shown in Figure \ref{pic_sed} with a red curve. The same characteristic shape in the mid-infrared is seen in hot dust-obscured galaxies (HotDOGs, \citealt{eise12, ster14a, asse15, pico15}), whose average SED from \citet{tsai15} is shown with a green curve. The remaining quasar templates in the SWIRE library all have lower infrared-to-optical ratios and a characteristic power-law shape which gives rise to the commonly used infrared color selection techniques for active galactic nuclei \citep{lacy04,ster05}. One of the typical power-law SEDs is shown with a magenta template in Figure \ref{pic_sed} (Mrk231 from \citealt{poll07}) demonstrating that such SEDs are inconsistent with the very steeply rising SEDs seen in our objects.

The apparent luminosities -- calculated as total fluxes measured between $\sim 0.1$\micron\ and $\sim 7$\micron\ in the rest-frame directly from SDSS, near-infrared and WISE data, multiplied by $4 \pi D_L^2$ under the assumption that the emission is isotropic -- are listed in Table \ref{tab:1}, with a median of $\sim 10^{46.8}$ erg s$^{-1}$. The template bolometric luminosities -- derived as the luminosities of the ``torus'' and HotDOG templates matched at the longest wavelength (in the W4 band) and also listed in Table \ref{tab:1} -- are higher yet, with a median of $10^{47.4}$ erg s$^{-1}$, primarily because the templates allow us to make a guess as to the shape of the infrared SED (between $\sim 7$\micron\ and $\sim 30$\micron) and include it even though most of the infrared bump in the SED is not covered by the data. Bolometric luminosity estimates derived from both ``torus'' and HotDOG templates are very similar to one another. They hinge on the shape of the SED in the far-infrared where in principle it could be contaminated by the star formation in the host galaxy, but \citet{tsai15} argue that it is not a significant fraction of the bolometric luminosity of the HotDOG template on the basis of molecular gas observations. If obscuration makes any of the near- and mid-infrared emission anisotropic, as is suspected to be the case in low-redshift type 2 quasars \citep{liu13b}, then the true bolometric luminosities of our targets are even higher. 

The extremely red quasars in our sample tend to have strong blue continua (necessary in order to be selected for follow-up spectroscopy in BOSS; \citealt{ross12}) with high-ionization emission lines being unambiguous signs of ionization by an active nucleus, whereas HotDOGs have fainter optical continua often consistent with being due to stars in the host galaxy \citep{asse15}. The median 2000\AA\ continuum luminosity of our four sources is $10^{45.1}$ erg s$^{-1}$. Using Starburst99 \citep{leit99} with calibration presented by \citet{igle04}, we find that this luminosity would require an unobscured star formation rate of $2.6\times 10^5 M_{\odot}$/year, which is not a plausible explanation for the ultra-violet continuum flux of our sources. Even though the ultra-violet emission of our sources is suppressed by a factor of 30-100 compared to that of type 1 quasars with matching mid-infrared luminosity (Fig. \ref{pic_sed}), it must be quasar-powered. 

Hamann et al. (in prep. 2016a) present an extensive discussion of the possible origin of the peculiar optical-to-infrared properties of extremely red quasars. Briefly, extinction by a large-scale ($\gg 1$ parsec) distribution of dust (including the toroidal obscuration of the classical unification model, \citealt{anto93}) is not a good explanation for the observed spectral energy distribution of our sources because of the strong blue continuum. The overall shape of the ultraviolet / optical SED can be obtained by reddening an SED of a type 1 quasar by $A_V\simeq 0.6-2.5$ mag, but even correcting for reddening the ultraviolet and optical power is inadequate for producing the observed mid-infrared via absorption and thermal re-emission. The observed infrared-to-optical ratios are too high even in the case of a quasar obscured over virtually all of the sky, with narrow holes through the obscuration and with the observer happening by chance to see through one of those openings toward the entire continuum source: with typical optical luminosities of $10^{45}$ erg s$^{-1}$ and bolometric corrections of $\la 10$ \citep{rich06}, the observed optical luminosities fall an order of magnitude short of powering the observed infrared emission.

One possibility for reconciling the discrepancy between the optical and the infrared continua is that the optical and ultraviolet spectra of our sources are reprocessed by scattering. Even if the direct view to the quasar is obscured, light may be able to propagate through openings in the obscuring material, scatter off the interstellar medium in the host galaxy on scales greater than obscuration and reach the observer. This mechanism is responsible for the high polarization of type 2 active galactic nuclei and quasars (e.g., \citealt{anto85, hine93, tran95a, youn96, zaka05, borg08}). In our sources the required scattering efficiency would be $1-10$\%, which is plausible and in line with values seen in low-redshift type 2 quasars \citep{zaka06, obie15}. In this scenario the continuum and the Balmer lines seen in the VLT spectra are produced close to the quasar, but are seen only in reflected light, whereas forbidden lines are produced in a different physical region, perhaps a more extended one, explaining the differing kinematics of forbidden and permitted emission lines. 

Another possibility is that the ultraviolet and optical continua are covered by a patchy screen of obscuring material, which lets only $\la 5\%$ of the light through, while largely preserving its spectral shape \citep{veil13b}. In order to provide partial covering of the ultraviolet continuum source which has sub-parsec size, the obscuring clouds must be individually even smaller and therefore are likely to reside at the same sub-parsec physical scales where the continuum emission is produced. Thus the patchy obscuration scenario is different from the large-scale obscuration scenario ruled out above in that in the patchy obscuration case the observer sees only a fraction of the continuum luminosity whereas in the narrow opening case the observer would see the entire continuum source, but only with a small probability. 

In both scenarios which are acceptable on the grounds of energetics -- patchy obscuration and scattering -- the equivalent widths of the emission lines produced close to the nucleus are expected to be unaffected. In the case of scattering, both the continuum emission and the line emission would be processed by a much more extended scatterer in the same way. In the case of patchy obscuration, the obscuring clouds must be present on the sub-parsec scale of the continuum emitter \citep{veil13b}, blocking ionizing photons that would otherwise be available for broad-line production and therefore again suppressing both the continuum and the line emission in the same way. The rest equivalent widths of H$\beta$ in our four sources (222, 86, 38 and 40\AA) are roughly consistent with values found in type 1 quasars with the highest [OIII] luminosities (REW[H$\beta$]$=96\pm 29$\AA\ for $L$[OIII]$>10^{43.0}$ erg s$^{-1}$, \citealt{shen11}), though the values in our sample show a large spread and their accuracy is $\sim 30\%$ because of difficulties in continuum placement. Thus, if the observed H$\beta$ emission arises close to the black hole and not in the extended region which produces the [OIII] emission, the observed equivalent widths are consistent both with patchy obscuration and with scattering. 

The REW of [OIII] seen in our objects (between 88 and 280\AA, with an additional lower confidence measurement of 480\AA\ in \oviii) are between those seen in type 1 quasars and type 2 quasars at the highest luminosity end \citep{gree14b}. In low-redshift type 2 quasars, obscuration suppresses the continuum while the bulk of the [OIII] emission originates outside obscuration, leading to very high equivalent width (in excess of 300\AA\ for $L$[OIII]$>10^{43.5}$ erg s$^{-1}$). In the extremely red quasars presented here, it is possible that obscuring material and [OIII]-emitting material are distributed on similar scales. In this case, the [OIII] line emission is also suppressed by obscuration, but to a lesser degree than the continuum. As a result, values of REW[OIII] that are intermediate between those seen in type 1s and type 2s might be naturally produced.

Assuming that underneath obscuration our sources are type 1 quasars with a typical spectral energy distribution and matching the observed infrared luminosities to the average quasar spectral energy distribution \citep{rich06}, we estimate their optical luminosities to be $\sim 10^{47}$ erg s$^{-1}$ or $M_{2500}=-28.2$ mag, placing them among the top $\sim$1\% most luminous quasars at this redshift \citep{rich06b}. Nonetheless, even at these extremely high luminosities unusually massive black holes are not required if the quasars are accreting at close to the Eddington limit ($1.3\times 10^{47}$ erg s$^{-1}$ for $10^9M_{\odot}$). While scattering and patchy obscuration might explain the overall shape of the SED (\citealt{asse16} reach similar conclusions for HotDOGs), neither mechanism naturally explains the shapes and extremely high equivalent widths of ultraviolet emission lines and the extreme kinematics of [OIII] in our sources, so additional elements are required to explain the phenomenology of extremely red quasars in a self-consistent manner (Hamann et al. in prep. 2016a).

\section{Discussion}
\label{sec:disc}

In the previous Section, we analyze rest-frame optical spectra and spectral energy distributions of four extremely red quasars. We measure fluxes and kinematics of the key rest-frame optical emission lines. In particular, we find very high velocity widths and blue-shifted asymmetries in the [OIII] emission. We measure [OIII]/H$\beta$ ratios and REW[OIII] values that are intermediate between those seen in type 1 and type 2 quasars. The SEDs show a steep rise from optical to the mid-infrared wavelengths, with a shape that is different from power-law SEDs seen in many active nuclei, with estimated bolometric luminosities $\sim 10^{47}$ erg s$^{-1}$. The four objects discussed in this paper demonstrate an unusual set of obscuration and outflow signatures. In this Section, we use the results of this analysis in a detailed comparison with quasars of other types (Section \ref{sec:type}) and in an estimate of the physical conditions and the energetics of these sources (Sections \ref{sec:energy}$-$\ref{sec:acc}).

\subsection{Comparison with other samples and type 1 / type 2 classification}
\label{sec:type}

\begin{figure*}
\centering
\includegraphics[scale=0.7, clip=true, trim=0cm 10cm 0cm 0cm]{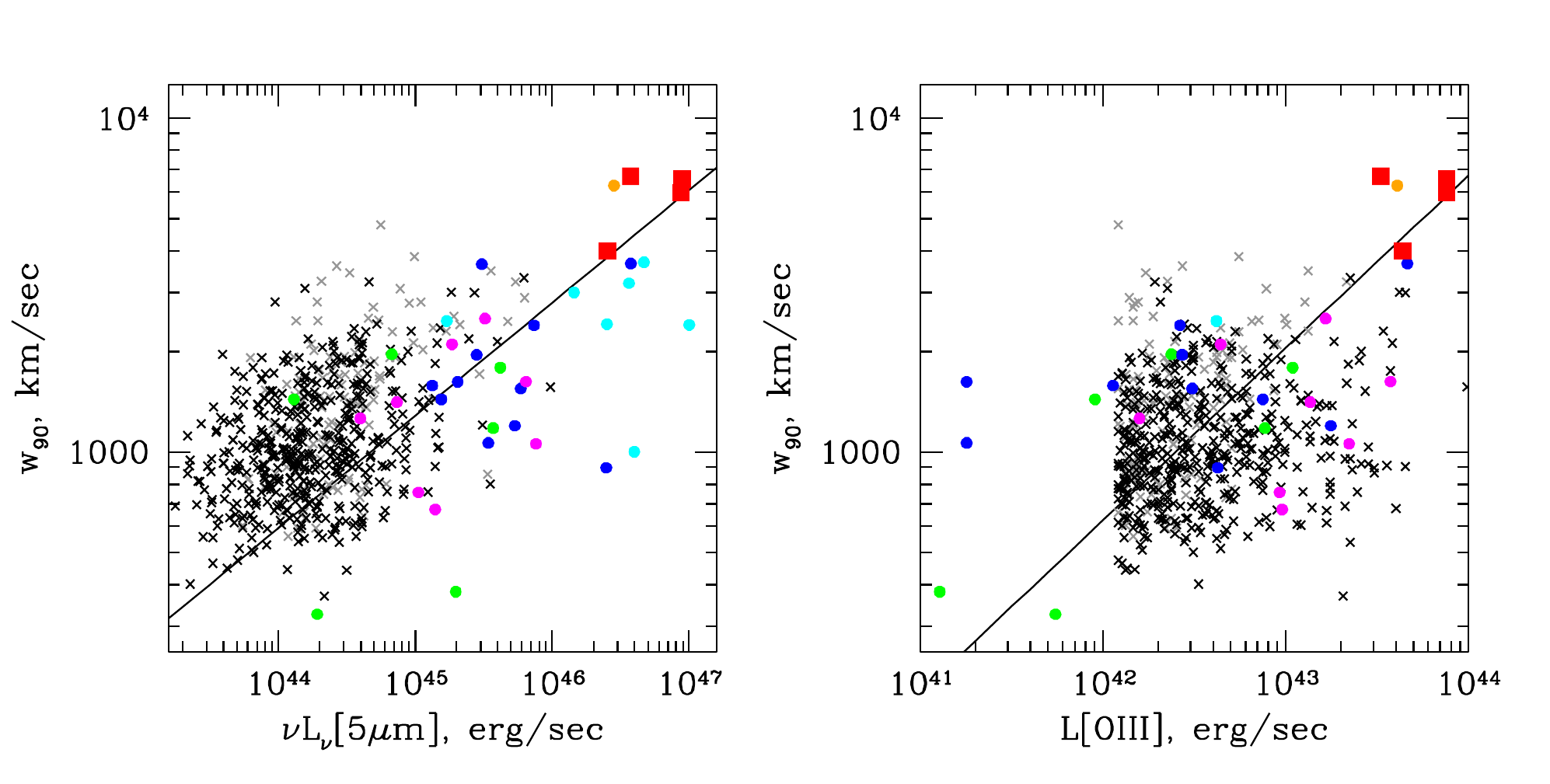}
\caption{[OIII] kinematics as a function of mid-infrared luminosities and [OIII] luminosities for the four objects presented in this paper (red squares, median $w_{90}=6270$ km s$^{-1}$) compared with those of red and type 2 quasar samples: $z<1$ type 2 quasars from \citet{zaka14} shown with grey and black crosses (black for peak signal-to-noise of [OIII] $\ga 20$, grey otherwise; median $w_{90}=1060$ km s$^{-1}$); X-ray selected obscured quasars at $z \sim 1.0-1.5$ from \citet{brus15} shown with green circles (median $w_{90}=1300$ km s$^{-1}$; infrared-selected red quasars at $z<1$ from \citet{urru12} shown with blue circles (fitting parameters for [OIII] emission for these objects are published by \citealt{brus15}; median $w_{90}=1580$ km s$^{-1}$); submm-selected $z\sim 2$ active galaxies from \citet{harr12} shown with magenta circles (median $w_{90}=1330$ km s$^{-1}$); and $z=3.4$ type 2 quasar SW022513 shown in orange from \citet{poll08,nesv11,poll11}. We also show a handful of type 1 quasars in cyan from \citet{cano12, komo15, carn15} wherever the data are available, with $w_{80}$ from \citet{carn15} estimated from their kinematic maps and scaled up to $w_{90}$ using the median value from \citet{zaka14}. The sharp cutoff for black points in the right panel is a selection effect, as only objects with $L$[OIII]$>10^{8.5}L_{\odot}$ were analyzed by \citet{zaka14}. The best-fit line is explained in Section \ref{sec:type}.}
\label{pic_extreme}
\end{figure*}

In Figure \ref{pic_extreme}, we compare the kinematics of the [OIII] in our objects against those of red and type 2 quasars. The [OIII] line widths seen in our sources are well above those seen in other samples. For example, there is only one type 2 quasar at $z<1$ in the sample of \citet{zaka14} with line kinematics comparable to those of \oxxii\ (which is the least kinematically disturbed of the four extremely red quasars studied here). The line widths in the \citet{zaka14} sample follow a roughly log-normal distribution (though with a tail at high widths) with $\log(w_{90},\mbox{km s}^{-1})=3.05\pm 0.19$ (sample mean and standard deviation). By this measure, the line widths of extremely red quasars presented here are 3$\sigma$ above the mean. Another source with comparable kinematics is a $z=3.5$ type 2 quasar from \citet{nesv11}. Figure \ref{pic_extreme} shows its most kinematically extreme central part as an orange dot, but the kinematics are more modest in the integrated spectrum of this object.

The strong correlation between [OIII] kinematics and infrared luminosity in type 2 quasars was already reported by \citet{zaka14}, and an increase in [OIII] widths and blueshifts with optical luminosity is also seen in type 1 quasars \citep{marz16, shen16}. Our new observations reinforce the idea that more infrared-luminous quasars tend to produce higher velocity outflows, as the four objects lie on the extreme end of the infrared luminosity / line width correlation. For Figure \ref{pic_extreme}, all rest-frame 5 \micron\ luminosities are calculated by power-law interpolating between closest adjascent mid-infrared flux measurements (\spi\ for objects from \citealt{brus15} and \citealt{harr12} and WISE for the rest).

To obtain the best-fit linear regressions shown in Figure \ref{pic_extreme} while accounting for the relative numbers of objects observed at different infrared luminosities, we conduct 100 trials in which we randomly draw four sources from the $z<1$ type 2 quasar population and four sources from the comparison samples, which we supplement with the four extremely red quasars. We fit a linear regression for each of the trials and we present the median of the trials in Figure \ref{pic_extreme}, with the best-fit relationships in the form:
\begin{eqnarray}
\log\left(\frac{w_{90}}{\mbox{ km s}^{-1}}\right)=-12.087+0.338\times \log\left(\frac{\nu L_{\nu}[5\micron]}{\mbox{ erg s}^{-1}}\right);\\
\log\left(\frac{w_{90}}{\mbox{ km s}^{-1}}\right)=-18.886+0.516\times \log\left(\frac{L{\rm [OIII]}}{\mbox{ erg s}^{-1}}\right).
\end{eqnarray}
The standard deviation in the slopes are $\simeq 0.1$ and $\simeq 0.2$, respectively (and the uncertainties in the intercept are strongly correlated with the uncertainties in the slope).

Traditionally, the observational distinction between type 1 (unobscured) and type 2 (obscured) active galaxies in the optical relied on the width of Balmer lines \citep{khac74}, with FWHM(H$\beta$)=1200 km s$^{-1}$ \citep{hao05a} being a well-motivated criterion developed for low-luminosity active nuclei. While outflows are common in low-luminosity active nuclei (e.g., \citealt{cren03}), they easily get stalled by the interaction with the interstellar medium and rarely result in kinematic disturbances of gas on larger scales, so [OIII] emission line profiles are dominated by gas moving in the galactic potential \citep{wils85, whit92b, nels96, gree05o3}. But active galaxies at higher luminosities often show deviations from galaxy-supported [OIII] kinematics. For this reason, \citet{zaka03}, \citet{reye08} and \citet{alex13} expanded the search for type 2 quasars to FWHM(H$\beta$)$\le 2000$ km s$^{-1}$. However, even with this higher cutoff some genuine type 2 quasars may be missed because forbidden lines alone -- and therefore the part of the H$\beta$ emission coming from the same physical region -- can be that broad \citep{nesv08,urru12,brus15}. The [OIII] line profiles presented here for four extremely red quasars constitute perhaps the most extreme known examples. 

As the [OIII] emission line widths enter squarely into the classical ``broad-line'' territory, the use of emission-line width for distinguishing type 2 and type 1 quasars becomes increasingly problematic and new approaches are necessary. One approach (employed in Section \ref{sec:oiii}) is to ask whether the forbidden and the permitted lines display the same kinematics \citep{reye08}. If they do, this might support the type 2 classification because the lines arise in the same physical region, so there is no contribution to the permitted lines from high-density regions (presumably close to the black hole). We find that [OIII] and H$\beta$ kinematics are inconsistent in three objects and consistent in one, perhaps suggesting that in the former three the vicinity of the black hole is directly observed whereas in the latter it is not. Unfortunately, this standard is ambiguous: [OIII] and H$\beta$ kinematics can be strikingly different in known type 2 quasars, especially in those with the strongest outflow signatures \citep{zaka14}, presumably due to the well-known but poorly understood stratification of the narrow-line region (e.g., \citealt{whit85c, dero86, veil91c, komo08b}). 

Another possibility is to use the classical emission-line diagnostic diagrams \citep{bald81, veil87}. Emission from the broad-line region would contribute to H$\alpha$ and H$\beta$, decreasing the apparent [OIII]/H$\beta$ and [NII]/H$\alpha$ ratios. None of our sources is close to the [OIII]$\lambda$5007\AA/H$\beta\sim 10$ value characteristic of the majority of type 2 quasars at low redshifts \citep{zaka14}, which would indicate an additional contribution to Balmer lines and thus a type 1 classification. Unfortunately, this method is also ambiguous, as the ionization conditions may be strongly modified by partial obscuration of the continuum and by the ram pressure of the winds \citep{ster15}, in which case low values of [OIII]/H$\beta$ can be naturally produced in the extended emission-line region. The rest equivalent widths of [OIII] which increase with stronger continuum obscuration are similarly ambiguous and are right in between type 1 and type 2 values \citep{gree14b}. 

\begin{figure}
\includegraphics[scale=0.7, clip=true, trim=0cm 10cm 10cm 0cm]{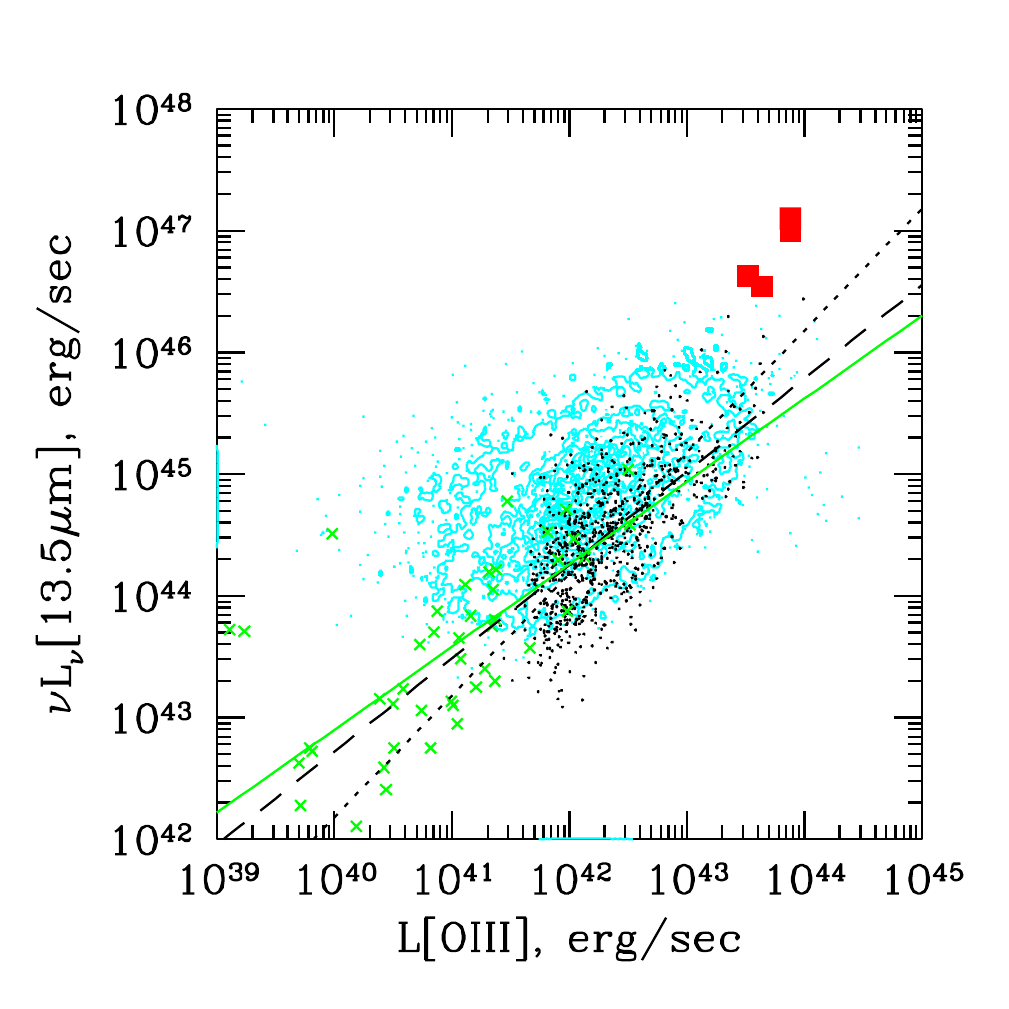}
\caption{[OIII] and mid-infrared (13.5 \micron) luminosities for the four sources in this paper (red squares; top right two sources overlap so closely that they have been artificially offset from each other for clarity), for type 2 quasars at $z<0.8$ from \citet{reye08} shown with black dots, for nearby Seyfert 2 galaxies from \citet{lama10} shown with green crosses, and for 17,000 type 1 quasars at $z<0.8$ from \citet{shen11} shown with cyan contours. The rest-frame 13.5 \micron\ luminosities of the four extremely red quasars in this paper are obtained from template fits shown in Figure \ref{pic_sed}. The dotted line fit to type 2 quasars and Seyfert 2s is made under the assumption that $\nu L_{\nu}[13.5 \micron] \propto L{\rm [OIII]}$, and we derive a coefficient of proportionality of $10^{2.18}$. An analogous relationship assumed for type 1 quasars yields a coefficient of proportionality of $10^{2.68}$, so they are 0.5 dex more infrared-luminous at fixed [OIII] luminosity. The solid green line is the power-law fit to Seyfert 2s from \citet{lama10}, with $\log \nu L_{\nu}$[13.5\micron]$=15.7+0.68\log L$[OIII]. The black dashed line is our power-law fit to both Seyfert 2s and type 2 quasars, $\log \nu L_{\nu}$[13.5\micron]$=12.1+0.77\log L$[OIII]}.
\label{pic_mir}
\end{figure}

In Figure \ref{pic_mir}, we compare the infrared and [OIII] luminosities for several samples of type 1 and type 2 active galaxies with those of our sources. For $z<0.8$ type 1 and type 2 quasars, the infrared luminosities are obtained by power-law interpolating between W3 and W4 measurements and correcting to the rest-frame 13.5\micron. For the extremely red quasars, there are no data beyond W4 (22\micron, corresponding to rest-frame 6.3\micron), and therefore the 13.5\micron\ luminosities are obtained by extrapolating from the longest wavelength observation using the fitted SED (Figure \ref{pic_sed}). 

[OIII] and infrared luminosities are strongly correlated in both type 1 and type 2 quasars, supporting the idea that both can serve as measures of quasar luminosity. At a given [OIII] luminosity, type 1 quasars are about three times more infrared luminous than type 2 quasars: $\nu L_{\nu}$[13.5 \micron]$=10^{2.68}L$[OIII] for type 1s, while the proportionality factor is $10^{2.18}$ for type 2s. Most likely this is due to the fact that even at 13.5 \micron, the thermal emission from quasars is not isotropic. This can be seen both in theoretical models of quasar obscuration \citep{pier92, frit06, honi06, scha08} and in SEDs of type 2 quasars which are notably redder than type 1 SEDs even in the mid-infrared \citep{liu13b}, though anisotropy of infrared emission can be strongly suppressed if the obscuring material is clumpy \citep{nenk08}.

The four extremely red quasars with median $\log(\nu L_{\nu}[13.5 \micron]/L{\rm [OIII]})=3.18$ lie at least an order of magnitude above the correlation for type 2 quasars and half an order of magnitude above the correlation for type 1s, suggesting that [OIII] is suppressed by extinction by a factor of 3$-$10 (we offer this comparison with a caution that the extremely red quasar sample is infrared-selected and thus may have an infrared-to-[OIII] ratio biased somewhat high). Furthermore, the strong effect of dust extinction on [OIII] is also clear from the shape of the emission lines: the large overall blueshift of [OIII] is likely produced due to the extinction of the redshifted part of the outflow by the observer-facing (blue-shifted) part of the outflow \citep{heck81, dero84, whit85a, wils85, cren10b}. Finally, the REW[OIII] values of extremely red quasars (intermediate between those of type 1 and type 2 quasars) also support the notion of dust extinction of the [OIII]-emitting region.

Another interesting possible connection is between our sources and the extreme end of the so-called Eigenvector 1 of the quasar population -- type 1 active galactic nuclei with strong FeII, low [OIII]/H$\beta$ and relatively strong [OIII] blueshifts which are exemplified in the low-redshift universe by the so-called narrow-line Seyfert 1 galaxies \citep{boro92, boro02, zama02}. Multiple lines of evidence suggest that Eigenvector 1 is driven by the increase in Eddington ratio \citep{wang96, laor97, will99} which we suspect is high in our targets based on their bolometric luminosity. At face value, our objects have neither strong FeII nor low [OIII]/H$\beta$ ratios, but that could be due to the suppression of the broad-line region by obscuration. However, both the blueshifts of [OIII] (extreme in our case, by comparison to those seen in low-redshift quasars, \citealt{boro05}) and the relatively low [OIII]/IR values may suggest some similarity with the Eigenvector 1 dominated objects. While this phenomenon is usually associated with relatively narrow H$\beta$ in the low-redshift universe, it is likely a luminosity effect \citep{sun15}: as we have targeted very high luminosity sources, we probe higher black hole masses ($\sim 10^9 M_{\odot}$) than those likely responsible for narrow-line Seyfert 1 objects ($\sim 10^7 M_{\odot}$) and therefore we find H$\beta$ widths to be in the normal range seen in type 1 quasars. 

The extremely red quasars in our sample present with a mix of type 1 / type 2 signatures which are likely due to a combination of patchy obscuration, scattering and extinction on nuclear scales and on the scales of the forbidden-line region. The standard line-width quasar type distinctions do not work in the regime of extreme outflow activity, and it is likely that the geometry and degree of obscuration in these sources are significantly different from the classical toroidal obscuration of the unification model \citep{anto93}. Perhaps ultimately all these properties may be explained in the context of dusty winds radiatively driven during accretion at a high Eddington ratio \citep{thom15, chan15}. 

\subsection{Energetics}
\label{sec:energy}

To calculate the energy of the outflow we need its mass and its velocity. We start by relating the observed velocity profiles with physical velocities of outflowing gas. We consider a spherically symmetric outflow with gas clouds moving radially away from the quasar with a distribution function $f(v_r){\rm d}v_r$, proportional to the luminosity emitted by gas with radial velocities between $v_r$ and $v_r+{\rm d}v_r$. The observed distribution of line-of-sight velocities $v_z$ is then
\begin{equation}
F(v_z)\propto \int_0^1\frac{{\rm d}a}{a}f\left(\frac{|v_z|}{a}\right), 
\label{eq_distrib}
\end{equation}
where integration over $a\equiv \cos\theta$ (where $\theta$ is the angle between a streamline and a line of sight) takes into account projection effects and the solid angle of the gas seen at a given $\theta$. The mean radial velocity of clouds within the outflow is $\langle v_r \rangle = \int_0^{\infty}v_rf(v_r){\rm d}v_r/\int_0^{\infty}f(v_r){\rm d}v_r$. Regardless of the distribution function $f(v_r)$, the mean physical velocity has a remarkably simple relationship with the observable line-of-sight value $v_z$: 
\begin{equation}
\langle v_r \rangle=2\langle |v_z| \rangle.
\label{eq_moments}
\end{equation}
Thus given an observed line profile, one can calculate $\langle |v_z| \rangle$ from observations and determine the average physical velocity of outflowing gas. 

The outflows in our extremely red quasars are affected by extinction and thus the spherically symmetric approximation is not directly applicable. To estimate $\langle |v_z| \rangle$ from our data, we assume that extinction mostly affects the redshifted part and use only the profiles at $v_z<0$. With this approach and using equation (\ref{eq_moments}), we estimate average physical velocities of the outflow of $\langle v_r \rangle\sim $3500, 4900, 2100 and 4500 km s$^{-1}$ in the four objects. Because obscuration more strongly affects the streamlines that are further away from the line of sight, these estimates are likely biased high. For example, the parts of the outflow propagating close to the plane of the sky which produce zero line-of-sight velocities could be suppressed by extinction of the observer-facing side of the outflow. As a result, in the discussion that follows we estimate that the physical outflow velocity is 3000 km s$^{-1}$ for all objects except \oxxii\ where the kinematics are less dramatic. 

A standard method of estimating the kinetic energy of the extended ionized gas is to use recombination lines to estimate the mass of the emitting hydrogen \citep{nesv06}. For this calculation, we would need the part of the H$\beta$ luminosity originating in the extended emission region, whereas our fits suggest that our H$\beta$ fluxes are dominated by emission much closer to the black hole (Section \ref{sec:type}), making a direct estimate of the extended gas mass difficult. Instead we assume that in the extended emission-line region the [OIII]/H$\beta$ ratio is close to its standard value of 10 \citep{dopi02}, which (together with an additional contribution to the H$\beta$ from the broad-line region) is consistent with the data. This allows us then to estimate the mass and kinetic energy of the outflowing gas:
\begin{eqnarray}
E_{\rm kin}=9.6\times 10^{58}\rm{erg}\times\nonumber\\
\left(\frac{\rm [OIII]/H\beta}{10}\right)^{-1}\left(\frac{L{\rm [OIII]}}{10^{10}L_{\odot}}\right)\frac{\langle v^2 \rangle}{(3000\rm{km\, s^{-1}})^2}\left(\frac{n_e}{100\rm{ cm}^{-3}}\right)^{-1}.
\label{eq_en}
\end{eqnarray}
No correction has been made here for the dust extinction of the [OIII] inferred from the [OIII]/IR ratio, [OIII] shape and the REW[OIII] values. If [OIII] is suppressed by a factor of 3$-$10, this energy estimate needs to be augmented by the same factor.

\subsection{Physical scale of the [OIII] emission}
\label{sec:scales}

If photo-ionization by the quasar is the dominant ionizing process, then narrow-line ratios \citep{dopi02} and equivalent widths \citep{bask05} suggest that most of the [OIII] emission originates in a region with ionization parameter $U\equiv \Phi/4\pi c r^2 n_H(r)=10^{-3}-10^{-2}$. If the ionization conditions producing forbidden emission lines in extremely red quasars are similar to those found in less extreme active galaxies, and if [OIII] is produced in regions with densities less than the critical density $n_H<8\times 10^5$ cm$^{-3}$, then we arrive at the minimal size of the [OIII]-emitting region \citep{hama11}:
\begin{eqnarray}
R=0.5{\rm kpc}\times \nonumber\\
\left(\frac{\nu L_{\nu}{\rm [1450 \AA]}}{10^{47}{\rm erg\, s^{-1}}}\right)^{1/2}\left(\frac{n_H}{8\times 10^5{\rm cm}^{-3}}\right)^{-1/2}\left(\frac{U}{0.003}\right)^{-1/2}.
\label{eq_size}
\end{eqnarray}
For comparison, the broad-line region for a black hole with mass $\sim 10^9 M_{\odot}$ is concentrated on scales $\la 1$ pc. Thus if the physics of [OIII] excitation is similar to that seen in nearby active galaxies, the high (suspected) intrinsic luminosity of extremely red quasars implies that the [OIII] emission must originate on scales comparable to the size of the host galaxy. For a more typical density of $10^3$ cm$^{-3}$ \citep{bask05}, the nominal emitting size is 14 kpc. Radiation-pressure-dominated models \citep{dopi02, ster14} suggest somewhat higher ionization parameters, between $U=0.01-0.1$, but even in this case the emitting size is between 2.5 and 8 kpc. 

These estimates, while very uncertain, are comparable to sizes of [OIII]-emitting nebulae directly measured in quasars at low redshifts with integral-field or long-slit spectroscopy \citep{liu13a, hain14} at the high luminosity end. In these observations, the surface brightness of [OIII] emission decreases away from the center, so much of the flux arises in the central few kpc, but the emission is detected out to $\ga 10$ kpc in some cases. Because of strong cosmological surface brightness dimming, resolving the sizes of extended line regions at high redshifts is very difficult, but existing measurements also suggest [OIII]-emitting sizes of a few kpc \citep{alex10, cano12, carn15, brus16}. Strong dust extinction can significantly reduce the estimated sizes: if the extinction within the line-emitting gas itself is high, then the effective luminosity of the quasar that enters equation (\ref{eq_size}) is suppressed and thus the size could be much smaller.

We can now combine the kinetic energy estimate from equation (\ref{eq_en}) with a size estimate to calculate the efficiency of conversion of the quasar luminosity into the kinetic energy of its wind. If $\tau$ is the characteristic time scale of the outflow (for example, $10^6$ years is the travel time to get to 3 kpc at 3000 km s$^{-1}$), then the efficiency is 
\begin{eqnarray}
\eta_{\rm wind}=\frac{E_{\rm kin}}{L_{\rm bol}\tau}=3\%\times \nonumber\\
\left(\frac{E_{\rm kin}}{10^{59}{\rm erg}}\right)\left(\frac{L_{\rm bol}}{10^{47}\rm{erg\, s^{-1}}}\right)^{-1}\left(\frac{v}{3000\rm{km\, s^{-1}}}\right)\left(\frac{R}{3{\rm kpc}}\right)^{-1}.
\end{eqnarray} 
Such efficiencies are in the ballpark of those required by galaxy formation models for suppressing excessive growth of massive galaxies \citep{hopk06}. This value of efficiency should be considered a lower limit since the observed [OIII] is likely strongly affected by extinction.

Normally shock ionization is energetically subdominant to photo-ionization in quasars \citep{ster15}. But in our objects, with their extreme kinematics, the possibility of shock ionization should be reconsidered in future analyses, especially with the indirect evidence for shocks supplied by strong [OIII]$\lambda$4363\AA\ (Section \ref{sec:4363}). 

\subsection{Acceleration mechanisms}
\label{sec:acc}

How can large-scale gas in a galaxy be accelerated to such high velocities? Relativistic jets are known to accelerate gas to velocities of up to a few thousand km s$^{-1}$. In a gas-rich medium, jets inflate an over-pressured cocoon \citep{bege89} which then expands into the surrounding medium and shocks and entrains the gas. In ionized gas observations similar to ours, the median FWHM of [OIII] seen by \citet{nesv06,nesv07,nesv08} in powerful high-redshift radio galaxies is $\sim 1300$ km s$^{-1}$, i.e., $2-3$ times smaller than that of extremely red quasars, but in the most extreme cases such as the central regions of MRC 0406$-$244 \citep{nesv08} the velocities can reach values comparable to those seen in our sample. The objects from these samples have a classical core-lobes radio morphology and a median radio luminosity of $\nu L_{\nu}$[1.4GHz]$=10^{44.5}$ erg s$^{-1}$, whereas ours are below $\nu L_{\nu}$[1.4GHz]$\sim 5\times 10^{41}$ erg s$^{-1}$. It is thus unlikely that our sources host relativistic jets of comparable power. Rather, their extremely high infrared luminosities and high inferred Eddington ratios suggest that the outflows are ultimately driven by the radiative energy release.

Acceleration by radiation pressure on dust is usually assumed to produce winds that move with velocities just above the escape velocity \citep{thom15}. As discussed in Section \ref{sec:scales}, we suspect that the [OIII] emission is produced on $\ga 1$ kpc scales, where the outflow velocities deduced from the [OIII] line shapes ($v\sim 3000$ km s$^{-1}$) are much greater than the local escape velocity (no more than a few hundred km s$^{-1}$). 

\citet{thom15} recently re-examined the case of acceleration by radiation pressure of dusty winds which are optically thick in the ultra-violet and concluded that much higher velocities, with a significant momentum boost, may be reached because the clouds or shells experience radiation pressure over the entire time they remain optically thick. We use their calculation of the terminal velocity of dusty winds, $v_{\infty}\simeq (4 R_{\rm UV}L/M_{\rm sh}c)^{1/2}$, where in our objects we estimate that $v_{\infty}\simeq 3000$ km s$^{-1}$, $L\simeq 10^{47}$ erg s$^{-1}$ is the luminosity of the source, $M_{\rm sh}$ is the mass of accelerated dusty shells and $R_{\rm UV}$ is the distance from the source where the shell becomes optically thin to ultra-violet photons. This distance is unknown in our objects, but we estimate it by remarking that [OIII] emission, despite originating on $\ga 1$ kpc scales, shows signs of significant obscuration, both in its shape (strong blue excess) and in the total luminosity (Section \ref{sec:type}). Therefore, we assume that the dusty, partially ionized clouds that produce [OIII] are still optically thick to ultra-violet emission at 1 kpc from the quasar and estimate that $R_{\rm UV}\ga 3$ kpc. Then for the mass of the shell, we find:
\begin{eqnarray}
M_{\rm sh}=6.9\times 10^8 M_{\odot}\times \nonumber\\
\left(\frac{R_{\rm UV}}{3{\rm kpc}}\right)\left(\frac{L}{10^{47}{\rm erg\, s^{-1}}}\right)\left(\frac{v_{\infty}}{3000{\rm km\, s^{-1}}}\right)^{-2},
\end{eqnarray}
its kinetic energy is $E_{\rm kin}=2R_{\rm UV}L/c=6.2\times 10^{58}$ erg and its momentum boost (the ratio of the momentum of the shell to the momentum of the absorbed photons) is $\dot{M}v_{\infty}c/L\simeq M_{\rm sh}v_{\infty}^2c/(R_{\rm UV}L)=4$. This kinetic energy is close to the estimate of the energy of the outflowing gas presented in Section \ref{sec:energy} which is based on the [OIII] luminosity, suggesting that dust acceleration is an energetically plausible mechanism in our case. 

An alternative possibility is that of light, ultra-fast ($v\ga 10,000$ km s$^{-1}$) radiatively-driven winds \citep{murr95, prog00, tomb15} which then collide with the interstellar medium of the surrounding host and shock and accelerate it to produce a galaxy-wide wind \citep{king11, fauc12b}, with the quasar photo-ionizing the resulting wind (perhaps with additional ionization contribution from shocks themselves). When the wind is first initiated by the quasar, the high inertia of the surrounding gas allows for a build-up of pressure within the central few tens of pc as the quasar continues depositing the energy into this region via the wind. When the pressure builds up to sufficiently high levels, the pressure force can overcome the inertia of the surrounding material and drive an energy-conserving flow. 

\section{Conclusions}
\label{sec:conc}

Extremely red quasars at $z\sim 2.5$ discussed in this paper were first identified by their red optical-to-infrared colors, while at the same time showing unambiguous signs of quasar activity in their optical (rest-frame ultra-violet) spectra such as strong CIV$\lambda$1549\AA\ \citep{ross15}. The infrared luminosities of these sources directly measured from WISE fluxes reach $10^{47}$ erg s$^{-1}$, placing them among the most luminous quasars known. These objects tend to display unusual ultra-violet emission line ratios -- with NV$\lambda$1240\AA/Ly$\alpha>1$ in some cases -- and stubby shapes of CIV$\lambda$1549\AA, lacking the extended wings characteristic of lines in ordinary quasars (Hamann et al. in prep. 2016a).

In this paper we present near-infrared (rest-frame optical) follow-up observations of four of these objects. We find extreme [OIII] kinematics in all 4 cases, where the [OIII]$\lambda\lambda$4959,5007\AA\AA\ lines are not only blended together, but are blended with H$\beta$ as well. Multiple kinematic measures are presented in Table \ref{tab:1}; for example, the FWHM of [OIII]$\lambda$5007\AA\ is between 2600 and 5000 km s$^{-1}$ and in 3 out of 4 objects, the line centroids are blueshifted relative to the H$\beta$ centroids by up to 1500 km s$^{-1}$. All four objects lie on the extreme end of the correlation between the [OIII] kinematics and the infrared luminosity (Figure \ref{pic_extreme}). The characteristic [OIII]-emitting gas velocities in our sources are 3$\sigma$ higher than the typical values seen in type 2 quasars at $z\la 1$ \citep{zaka14}. 

At these extreme velocities, the gas cannot be confined by any realistic galaxy potential and is thus likely to escape from the galaxy; we suggest that these objects may be signposts of the extreme ``blow-out'' phase of quasar feedback observed at the peak epoch of galaxy formation. Our energetics estimates suggest that at least a few per cent of the bolometric luminosity is converted into the kinetic energy of the outflowing ionized gas. Such feedback efficiency is in the range of values required from quasar feedback by galaxy formation models \citep{hopk06, choi12}. The observed winds are dusty, as manifested by the strong line asymmetry and by an overall deficit of [OIII] relative to the amount expected based on the infrared luminosity. Although we cannot spatially resolve the outflows with our current data, we estimate that the [OIII]-emitting gas must be extended on scales of $\ga 3$ kpc and therefore the wind is progressing on galaxy-wide scales. 

Although the infrared-to-optical colors suggest significant levels of obscuration, their spectral energy distributions are inconsistent with extinction by a foreground screen of dust, and the ultra-violet continua of these objects ($\nu L_{\nu}\sim 10^{45}$ erg s$^{-1}$) are too strong to be due to the host galaxy. The most likely explanation is that we observe the continuum source through patchy obscuration or reprocessed by scattering in the interstellar medium of the host galaxy or perhaps in the dusty outflow itself. 

Hamann et al. (in prep., 2016a) demonstrate that extremely red quasars such as the ones we study here are likely more common than are hot dust-obscured galaxies (HotDOGs). HotDOGs, in turn, appear to be as common as type 1 quasars of the same bolometric luminosity \citep{asse15}. If our subsequent observations demonstrate that the [OIII] kinematics we see in these four sources are common for the extremely red quasars, this would imply that the ``blow-out'' phase may be a common phase in evolution of extremely luminous quasars. Whether every massive galaxy undergoes a period of such luminous quasar activity is not yet known. But it is clear that with observed velocities and energetics of gas removal, such winds could have a profound impact on the evolution of galaxies in which they occur, in a direct manifestation of quasar feedback on their hosts. 

\acknowledgments

NLZ is grateful to C.-A. Faucher-Gigu\`ere, J. Krolik, J. Stern, T. Urrutia and the anonymous referee for discussions and suggestions that helped improve the paper, and to the Institute for Advanced Study for continued hospitality.  

\bibliographystyle{apj}

\input{paper.bbl}
\clearpage
\begin{deluxetable}{l|l|l|l|l}
\tabletypesize{\small}
\tablecolumns{5}
\tablecaption{Emission line parameters\label{tab:1}}
\startdata
\input{table1.dat.edited}
\enddata
\tablecomments{Notes: \\
(b) The H$\beta$+[OIII] fits listed here are based on the kinematically untied model of one Gaussian for H$\beta$ and two Gaussians for [OIII]. In \oxxii, both kinematically tied and kinematically untied models are acceptable and the H$\beta$ flux is given for both (lower value for the kinematically tied fit).\\
(c) In the H$\alpha$+[NII] fits, H$\alpha$/H$\beta$ are fixed to Case B value of 2.85 in \oviii, \oxii, and \oxxiii, and therefore the line ratios and the equivalent widths are given in parentheses. The best fits with the same kinematics, but without the restriction on the amplitude, have H$\alpha$ fluxes lower by 50\%, 19\% and 1\%, respectively. In \oviii\ H$\alpha$ falls into the last order where the flux calibration is very uncertain, so H$\alpha$ flux should be considered an estimate. In \oxxii, H$\alpha$/H$\beta$ ratio depends on whether the kinematically tied fit (higher value) or the kinematically untied fit (lower value) is adopted for H$\beta$+[OIII], and [NII] is not detected because of the low signal-to-noise ratio of the longest wavelength order in this object. \\
(d) Most kinematic parameters are given to more than necessary number of significant digits. The uncertainties in line fits are dominated by systematics such as continuum placement (including possible FeII contamination) and telluric corrections. Performing the fits with and without the telluric corrections and with and without the FeII contribution, we estimate that in single-Gaussian fits the line width (and consequently line luminosity) uncertainties are $\la$ 15\% and those in the centroid velocities are $\la 200$ km s$^{-1}$. In double-Gaussian fits, the individual components can change more signifantly, but these uncertainties still apply for the non-parametric widths and median velocities. \\ 
(e) In \oxxii, H$\beta$ luminosity is given for both kinematically tied (lower value) and kinematically untied (higher value) models. The value of rest equivalent width of [OIII] in \oviii\ is poorly determined because the continuum is strongly affected by telluric absorption between $H$ and $K$ bands. All H$\alpha$ equivalent widths are tentative because the continuum is poorly constrained. Apparent luminosities are obtained by integrating the flux over the range probed by direct observations, from SDSS to WISE (Figure \ref{pic_sed}). 13.5\micron\ luminosities are obtained by extrapolating from WISE bands using the torus and the HotDOG templates. Bolometric luminosities estimated by fitting torus and HotDog templates are also listed.}
\end{deluxetable}

\end{document}

%% file: table1.dat.edited
object name & \oviii   & \oxii & \oxxii  & \oxxiii \\
\hline
(a) Basic information: & & & & \\
object R.A. & 08:34:48.48 & 12:32:41.73 & 22:15:24.00 & 23:23:26.17 \\ 
object declination & $+$01:59:21.1 & $+$09:12:09.3 & $-$00:56:43.8 & $-$01:00:33.1\\
$r$-band AB magnitude & 21.20$\pm$0.05 & 21.11$\pm$ 0.05 & 22.27$\pm$0.12 & 21.62$\pm$ 0.08 \\ 
initial (pipeline) redshift $z_{\rm in}$ &  2.591 &  2.381 &  2.509 &  2.356 \\ 
\hline
(b) Results of H$\beta$+{[}OIII{]}$\lambda\lambda$4959,5007\AA\AA\ fitting: & & & & \\
H$\beta$ centroid relative to $z_{\rm in}$ frame, km s$^{-1}$ & 54 & 1036 & $-$1465 & 1257 \\ 
H$\beta$ velocity dispersion, km s$^{-1}$ & 2103 & 2119 & 2128 & 1564 \\
H$\beta$ flux, in $10^{-17}$ erg sec$^{-1}$ cm$^{-2}$ & 63.2 & 56.2 & (15.5$-$23.1) & 32.3 \\
{[}OIII{]} Gaussian component 1: & & & & \\
peak $F_{\lambda}$, $10^{-17}$ erg s$^{-1}$/cm$^2$/\AA & 0.444 & 0.432 & 0.413 & 0.322 \\ 
central velocity rel. to H$\beta$, km s$^{-1}$ &  $-$828 &  $-$150 &  $-$123 &    11 \\ 
velocity dispersion, km s$^{-1}$ &  1872 &   951 &  1259 &   957 \\ 
{[}OIII{]} Gaussian component 2: & & & & \\
peak $F_{\lambda}$, $10^{-17}$ erg s$^{-1}$/cm$^2$/\AA & 0.236 & 0.412 & 0.211 & 0.123 \\ 
central velocity rel. to H$\beta$, km s$^{-1}$ & $-$1155 & $-$2643 &  1140 & $-$3060 \\ 
velocity dispersion, km s$^{-1}$ &   381 &  1964 &   315 &  1903 \\ 
{[}OIII{]}$\lambda$5007\AA\ flux, in $10^{-17}$ erg sec$^{-1}$ cm$^{-2}$ & 138.6 & 172.7 & 86.1 & 76.1 \\
\hline
(c) Results of H$\alpha$+{[}NII{]}$\lambda\lambda$6548,6583\AA\AA\ fitting: & & & & \\
H$\alpha$/H$\beta$ & (2.85) & (2.85) & (11.37$-$7.63) & (2.85) \\
{[}NII{]}6583\AA/H$\alpha$ & (0.671) & (0.333) & n/a & (1.077) \\
{[}NII{]} central velocity rel. to H$\beta$, km s$^{-1}$ & 292 & $-$115 & n/a & 238 \\
{[}NII{]} velocity dispersion, km s$^{-1}$ & 599 & 509 & n/a & 660 \\
\hline
(d) Kinematic measurements of [OIII]: & & & & \\
$v_{50}$, median vel. of {[}OIII{]} rel. to H$\beta$, km s$^{-1}$ &  $-$937 & $-$1520 &    79 &  $-$769 \\ 
$w_{80}$ velocity width of {[}OIII{]}, km s$^{-1}$ &  4576 &  5311 &  3554 &  5409 \\ 
$w_{90}$ velocity width of {[}OIII{]}, km s$^{-1}$ &  5970 &  6571 &  4005 &  6690 \\ 
FWQM[OIII]: full width at quarter max., km s$^{-1}$ &  5213 &  6785 &  4051 &  6172 \\ 
FWHM[OIII]: full width at half max., km s$^{-1}$ &  2811 &  4971 &  3057 &  2625 \\
\hline
(e) Luminosities and equivalent widths: & & & & \\
log ($L${[}OIII{]} in erg s$^{-1}$) & 43.88 & 43.88 & 43.64 & 43.52 \\ 
log ($L$[H$\beta$] in erg s$^{-1}$) & 43.54 & 43.40 & (42.89$-$43.07) & 43.14 \\ 
log ($L_{\rm apparent}$ in erg s$^{-1}$) & 46.9 & 47.0 & 46.5 & 46.7 \\
log ($\nu L_{\nu}$ at 13.5\micron\ in erg s$^{-1}$) & 47.1 & 47.0 & 46.5 & 46.6 \\
log ($L_{\rm torus template}$ in erg s$^{-1}$) & 47.6 & 47.6 & 47.1 & 47.2 \\
log ($L_{\rm HotDOG template}$ in erg s$^{-1}$) & 47.6 & 47.5 & 47.0 & 47.1 \\
rest eq. width of [OIII] (in \AA) & (485) & 280 & 213 & 88 \\
rest eq. width of H$\beta$ (in \AA) & 222 & 86 & 38 & 40 \\
rest eq. width of H$\alpha$ (in \AA) & (425) & (180) & (524) & (71) 